\documentclass[useAMS,usenatbib,usegraphicx]{mn2e}

\newcommand{\msun}{M_\odot}

\newcommand{\angstrom}{{\rm \AA}}
\newcommand{\Teff}{$T_{\rm eff}$}
\newcommand{\logg}{$\log g$}

\def \aj {AJ}
\def \mnras {MNRAS}
\def \apj {ApJ}

\def \aap {A\&A}

\def \pasp {PASP}

\usepackage{amssymb}
\usepackage{lscape}

\title[Spectral models: solar and $\alpha$-enhanced]
 {Spectral models for solar-scaled and $\alpha$-enhanced stellar populations}
\author[P. Coelho et al.]
{P.~Coelho,$^{1,3,4}$\thanks{E-mail: pcoelho@iap.fr (PC), bruzual@cida.ve (GB), charlot@iap.fr (SC), 
weiss@mpa-garching.mpg.de (AW), barbuy@astro.iag.usp.br (BB), jason.ferguson@wichita.edu (JF)}
G.~Bruzual,$^2$
S.~Charlot,$^1$ 
A.~Weiss,$^3$
B.~Barbuy$^4$ 
and J.W.~Ferguson$^5$\\
$^1$ Institut d'Astrophysique, CNRS, Universit\'e Pierre et Marie Curie, 98 bis Bd Arago, 75014 Paris, France\\
$^2$ Centro de Investigaciones de Astronomia (CIDA), Merida, Venezuela \\
$^3$ Max-Planck-Institute f\"ur Astrophysik, Karl-Schwarzschild-Strasse 1, 85740 
Garching bei M\"unchen, Germany \\
$^4$ Universidade de S\~ao Paulo, IAG, Rua do Mat\~ao 1226,
 S\~ao Paulo 05508-900, Brazil\\
$^5$ Department of Physics, Wichita State University, Wichita, KS 67260-0032, 
USA}
\date{Accepted 2007 August 13. Received 2007 August 2; in original form 2007 April 19}

\pagerange{\pageref{firstpage}--\pageref{lastpage}} \pubyear{2006}

\def\LaTeX{L\kern-.36em\raise.3ex\hbox{a}\kern-.15em
 T\kern-.1667em\lower.7ex\hbox{E}\kern-.125emX}

\begin{document}

\label{firstpage}

\maketitle

\begin{abstract}
We present the first models allowing one to explore in a consistent way the
influence of changes in the $\alpha$-element-to-iron abundance ratio on the high-resolution
spectral properties of evolving stellar populations. The models cover the wavelength range 
from 3000~{\AA} to 1.34$\,\mu$m at a constant resolution of FWHM=1~{\AA} and a sampling
of 0.2 $\angstrom$, for overall metallicities in the range $0.005\leq Z\leq 0.048$ and for
stellar population ages between 3 and 14 Gyr. These models are based on a recent library
of synthetic stellar spectra and a new library of stellar evolutionary tracks, both computed
for three different iron abundances ([Fe/H] = $-$0.5, 0.0 and 0.2) and two different 
$\alpha$-element-to-iron abundance ratios ([$\alpha$/Fe] = 0.0 and 0.4). 
We expect our fully synthetic models to be primarily useful for evaluating 
the differential effect of changes in the $\alpha$/Fe ratio on spectral 
properties such as broad-band colours and narrow spectral features. In addition,
we assess the accuracy of absolute model predictions in two ways: first,
by comparing the predictions of models for scaled-solar metal abundances 
([$\alpha$/Fe] = 0.0) to those of existing models based on libraries of observed stellar
spectra; and secondly, by comparing the predictions of models for $\alpha$-enhanced metal 
abundances ([$\alpha$/Fe] = 0.4) to observed spectra of massive early-type galaxies in the Sloan
Digital Sky Survey Data Release 4. We find that our models predict accurate
strengths for those spectral indices that are strongly sensitive to the abundances
of Fe and $\alpha$ elements. The predictions are less reliable for the strengths of other spectral
features, such as those dominated by the abundances of C and N, as expected from the fact that the
models do not yet allow one to explore the influence of these elements in an independent way. We
conclude that our models are a powerful tool for extracting new information about the chemical
properties of galaxies for which high-quality spectra have been gathered by modern surveys.

\end{abstract}


\section{Introduction}

The chemical abundance patterns in the spectra of stellar populations are direct
tracers of the histories of star formation and chemical enrichment of galaxies. 
These patterns are sensitive to 
the different relative amounts of metals produced on short time-scale 
($\sim10^7$ yr) by Type II supernovae and on longer time-scale (a few x$10^9$ yr) by 
Type Ia supernovae. 
In recent years, modern spectroscopic galaxy surveys have gathered 
hundreds of thousands of high-quality spectra, in which such signatures can be 
analyzed \citep[e.g.][]{2dFGRS,sdss-dr4}. We must be able to interpret these
spectral signatures to constrain the past histories of star formation and chemical
enrichment of the galaxies.

Evolutionary stellar population synthesis, i.e. the modelling of the spectral 
energy distribution (SED) emitted by evolving stellar populations, is a natural 
approach to studying the stellar content of galaxies. A current limitation in this area is that
all models of high-resolution galaxy spectra rely on the spectral properties 
of stars in our own Galaxy. This implies that the models are tuned to the specific 
chemistry and star formation history of the Milky Way.  This is restraining, 
because stars in external galaxies are known to exhibit other
abundance patterns. For example, in massive early-type galaxies, light $\alpha$ 
elements (O, Ne, Mg, Si, S, Ca and Ti, produced mainly by Type II supernovae) 
are typically more abundant relative to iron (produced mainly by Type Ia 
supernovae) than in the Milky Way disk \citep[e.g.][]{wfg92}. Hence, widely used
population synthesis models such as those developed by \citet{vazdekis99}, 
\citet{BC03} and \citet{PEGASE-HR}, should be applied with caution to the
analysis of high-resolution spectra of external galaxies. 

Pioneer population synthesis models have been developed to quantify the 
influence of changes in metal abundance ratios on the strengths of `Lick/IDS 
indices' that measure a number of strong absorption features in galaxy 
spectra \citep[see][]{burstein+84,worthey+04,trager+98}. These models rely on a 
semi-empirical approach, which consists in applying theoretical response functions
to empirical fitting functions in order to predict the strengths of Lick/IDS indices with
variable abundance ratios \citep[e.g.,][]{trager+00a,TMB03,proctor+04,tantalo04,
lee+worthey05,annibali+07,schiavon07}.  The empirical {\it fitting functions} describe how
spectral indices vary as a function of the stellar parameters effective temperature
\Teff, gravity \logg, metallicity [Fe/H] and, sometimes, abundance [X/Fe] of 
another element relative to iron \citep[e.g.][]{diaz+89,jorgensen+92,worthey+04,
borges+95,schiavon07}. Theoretical {\it response functions} \citep[e.g.][]{TB95,
houdashelt+02,korn+05} quantify how the indices change due to variations of
individual chemical elements. Typically, these response functions are calculated 
only for three combinations of \Teff\ and \logg, which correspond to a 
main-sequence (MS) star, a turn-off star and a red-giant star for a 5 Gyr-old stellar 
population. All other stars of a same evolutionary phase are assumed to respond 
in the same way to abundance variations. Alternatively, theoretical fitting 
functions with explicit dependence on $\alpha$/Fe \citep{barbuy+03} can be 
employed to compute $\alpha$-enhanced models \citep{oliveira+05}. By 
construction, all these models are limited to the study of a few selected 
spectral indices.

Recent progress in the modelling of high-resolution stellar spectra is opening the
door to a new kind of models, in which the effects of abundance variations can be
studied at any wavelength. This is enabled by the publication of several 
libraries of theoretical, high-resolution stellar spectra for both solar-scaled
and $\alpha$-enhanced chemical mixtures \citep{MARCS05,phoenix05,
coelho+05,malagnini+05,munari+05}. Libraries with solar-scaled
mixtures have already been used to construct high-resolution population synthesis 
relying purely on theoretical spectra \citep[e.g.][]{delgado+05,zhang+05}. Such spectra
are easier to handle than observed spectra because they have arbitrary wavelength 
resolution, infinite signal-to-noise ratio (S/N), and they correspond to well-defined
stellar parameters. The disadvantage of synthetic spectra is that they are limited by the
uncertainties inherent in the underlying model atmosphere calculations (such as the
neglect of departures from local thermodynamical equilibrium and of chromospheric, 
three-dimensional and sphericity effects) and by potential inaccuracies in the parameters
of the atomic and molecular transitions line lists (line wavelengths, energy levels, 
oscillator strengths and damping parameters; see e.g. \citealt{kurucz06}). Despite these limitations, theoretical model spectra represent
a unique opportunity to study the response of any spectral feature to abundance 
variations in galaxy spectra. For such a study to be meaningful, one must also include
the effects of abundance variations on the evolution of the stars, which cannot be
neglected for metallicities larger than about half solar \citep{wpm:95,sw:98,vsria:2000,
sgwc:2000}. 

In this paper, we combine the library of synthetic stellar spectra recently 
published by \citet[][ hereafter C05]{coelho+05} with a new library of stellar
evolutionary tracks computed by \citet{weiss+07}, to predict the high-resolution 
spectral properties of evolving stellar populations with different metallicities and
$\alpha$/Fe abundance ratios. For the purpose of this study, we extend the range of 
stellar effective temperatures and improve the spectrophotometric calibration of the
C05 library (see Sections~\ref{c05ext}--\ref{fluxcalib}). A salient feature of our 
model is that both the library of synthetic stellar spectra and that of stellar 
evolutionary tracks are computed for the same mixtures of chemical elements, listed
in Table~\ref{tab_model_param}. At fixed [Fe/H], we assume that in the
`$\alpha$-enhanced' model, the abundances of all classical $\alpha$ elements, i.e.
O, Ne, Mg, Si, S, Ca and Ti, are increased by 0.4\,dex relative to the scaled-solar
model.\footnote{There is no consensus in the literature on whether it is more 
appropriate to fix the iron abundance [Fe/H] or the total metal abundance [Z/H] when
exploring models with different $\alpha$/Fe \citep[see e.g. the different
methods in][]{PS02}.
C05, following \cite{atlasodfnew}, fix [Fe/H] when computing theoretical stellar 
spectra for different  $\alpha$/Fe. We adopt the same convention, which is motivated
by the fact that in high resolution stellar spectroscopic studies, [Fe/H] can be 
directly derived from the equivalent widths of Fe lines. The total metallicity [Z/H] 
is a less accurate value, being computed based on assumptions for those
elements whose abundances could not be derived.} 
We adopt the \citet[ hereafter BC03]{BC03} code to compute the spectral 
evolution of stellar populations based on these libraries of stellar spectra and 
evolutionary tracks. The resulting models cover the wavelength range from 3000~{\AA}
to 1.34 $\,\mu$m at a constant resolution of FWHM=1~{\AA} and a sampling of 
0.2 $\angstrom$, for overall metallicities in the range $0.005\leq Z\leq 0.048$ and for
stellar population ages between 3 and 14 Gyr.

The paper is organized as follows. In Sections~\ref{stelevol} and \ref{stelsed} below,
we describe in detail the main two ingredients of our models: the library of stellar 
evolutionary tracks and the library of synthetic stellar spectra, respectively. 
Section~\ref{popsyn} provides an overview of the stellar population synthesis code. In
Section~\ref{synsed}, we make a first assessment of the performance of our fully synthetic
models by comparing the predictions for scaled-solar stellar populations to those of 
models based on libraries of observed stellar spectra. Then, in Section~\ref{modpred},
we explore the effects of the $\alpha$ enhancement on the spectroscopic and photometric 
properties of stellar populations. We show that the spectra of $\alpha$-enhanced models 
compare well to observed spectra of massive early-type galaxies in the Sloan Digital Sky 
Survey Data Release 4 (SDSS-DR4). We also compare our predictions for Lick-index strengths
for different $\alpha$/Fe with respect to those of previous models. Section~\ref{summary} summarizes 
our results.


\begin{table}
\caption{Chemical composition of the models}
\label{tab_model_param}
\begin{center}
\begin{tabular}{cccccccc}
\hline
Label & X & Y & Z & [Fe/H] & [$\alpha$/Fe] & [Z/Z$_\odot$] \\
\hline
$m05p00$ & 0.743 & 0.251 & 0.005 & $-$0.5 & 0.0 & $-$0.5 \\
$m05p04$ & 0.739 & 0.250 & 0.011 & $-$0.5 & 0.4 & $-$0.2 \\
$p00p00$ & 0.718 & 0.266 & 0.017 &  0.0 & 0.0 & 0.0  \\
$p00p04$ & 0.679 & 0.289 & 0.032 &  0.0 & 0.4 & 0.3  \\
$p02p00$ & 0.708 & 0.266 & 0.026 &  0.2 & 0.0 & 0.2  \\
$p02p04$ & 0.642 & 0.310 & 0.048 &  0.2 & 0.4 & 0.5  \\
\hline
\end{tabular}
\end{center}
\end{table}

\section{Stellar evolution models}\label{stelevol}

Above approximately half solar metallicity, the
$\alpha$ enhancement changes the evolutionary time-scales and
isochrones in non-negligible ways
\citep{wpm:95,sw:98,vsria:2000,sgwc:2000}. At fixed total metallicity,
$\alpha$-enhanced isochrones tend to be warmer and slightly fainter
than solar-scaled ones, while at fixed [Fe/H] the increased total
metallicity of the $\alpha$-enhanced models leads to cooler colours
and fainter brightness.  It is crucial
therefore that single stellar population (SSP) models include tracks computed with proper
abundance patterns. And it is important that the assumptions for the abundance 
pattern are consistent among the ingredients.

\subsection{Stellar evolution code}

The stellar evolution models for this project were calculated using the Garching Stellar
Evolution Code, which was described in detail in
\cite{wsch:2000}. We use the OPAL equation of state \citep{rsi:96},
nuclear reaction rates mainly from \cite{CFHZ:85}, \cite{CF:88} and
\cite{adel:98}, neutrino emission rates from \cite{ihnk:96} and
\cite{hrw:94}. We use standard mixing length theory with a value of
$\alpha_\mathrm{MLT} = 1.6$, which is obtained from a solar model
calibration. We ignore any effects of overshooting and semi convection, and
diffusion is included neither in the present models nor
in the solar one. The models are evolved at constant mass. They are
therefore the most canonical ones possible. Inclusion of any of the
effects mentioned above would introduce additional parameters and
assumptions. Since the main purpose of this paper is to investigate
the influence of $\alpha$-element abundances, we thought it more
reasonable to keep the number of assumptions as low as possible.

Element variations enter stellar models at various places, and we have
taken great care to consider these as consistently as possible. The
most obvious instance is the chemical mixture for the initial
model. We start from the solar composition of \cite{gs:98}, adding a
constant enhancement of 0.4~dex to O, Ne, Mg, Si, S, Ca and Ti. The
final abundances are given in \citet[ table~1]{weiss+07}. The molecular
weight is calculated from these abundances. Although one would like to
take the detailed abundances into account also in the equation of
state, this is no longer trivial. 
In this respect, our models are not
self-consistent, as we consider only the total metallicity (and the
hydrogen and helium abundances) in the calculation of the equation of
state. It is
generally supposed that this is a sufficiently accurate approximation.

In the nuclear network, element abundances enter explicitly the
reaction rates and therefore can be taken into account
straightforwardly. 

Finally, and most importantly for the stellar effective temperature and
therefore observed colours, spectra and line indices, the
opacities have to be calculated as accurate as possible.  Throughout
this paper we are using Rosseland opacity tables from the
OPAL-website\footnote{\tt http://www-phys.llnl.gov/Research/OPAL}
\citep{ir:96} for high temperatures together with low-temperature
molecular opacity tables calculated by one of us (JF), employing the
code by \cite{fa:2005}. The most important feature of the final
opacity tables is that they were produced for exactly the same element
composition, both in the solar and $\alpha$-enhanced case.  Details on
the implementation of opacity tables in our code can be found in
\cite{weiss+07}.

We initially intended to use the set of $\alpha$-enhanced tables
already available to us, which include molecular opacities by
\cite{af:94}. These data were used, for example, by \cite{sgwc:2000},
and are for a slightly different element
composition of varying $\alpha$ enhancement factors. We assumed initially that
the internal table composition differences would not affect the models
significantly. However, as discussed in \cite{weiss+07}, this
assumption proved to be wrong, such that we produced completely
consistent and up-to-date new tables. For the effect of
$\alpha$-element variations on the models and isochrones, see below.

\subsection{Models}

We followed the evolution of low-mass stars ($M \le 2\, \msun$) from
the MS up to the core helium flash, that of intermediate
mass stars ($3 \le M/\msun \le 7$) into the early asymptotic giant
branch (AGB), and that of $M=10\,\msun$ to the beginning of core carbon burning. 

The mass grid in detail was: $M  = 0.60, \ldots (0.05) \ldots,$
1.20, 1.50, 2.00, 3.00, 5.00, 7.00, 10.00\,$\msun$. 
The various compositions are given in Table~1.

\begin{figure}	
\includegraphics[scale=0.5]{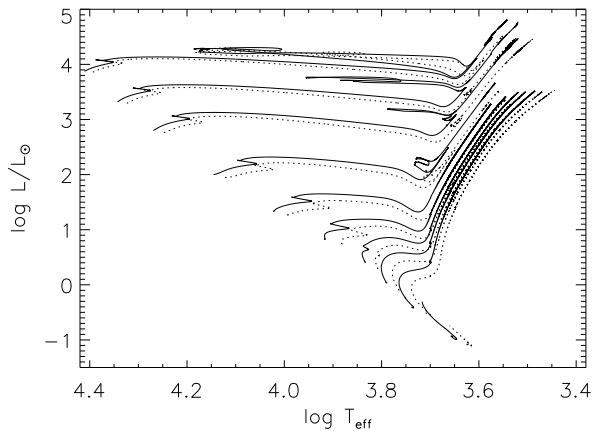}
\caption{Evolution of models at $\mathrm{[Fe/H]}=-0.5$ with mixture
[$\alpha$/Fe] = 0.0 ($m05p00$ in Table 1; solid lines) and [$\alpha$/Fe] = 0.4 
($m04p04$;
dotted lines). For clarity, only tracks for $M/\msun =$ 0.6, 0.8, 1.0,
1.2, 1.5, 2.0, 3.0, 5.0, 7.0, and 10.0 are shown.}  
\protect\label{fa:1}
\end{figure}

\begin{figure} 
\includegraphics[scale=0.5]{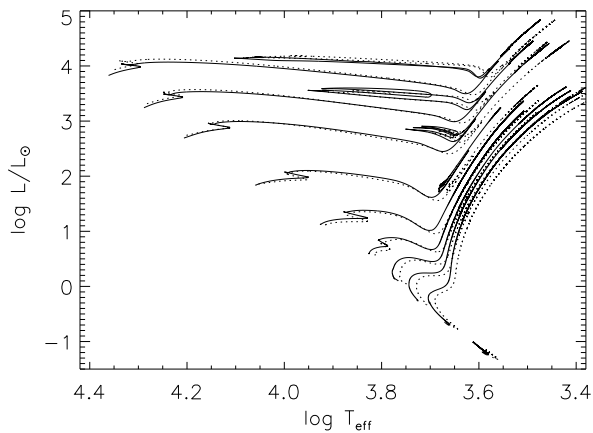}
\caption{As Fig.~\ref{fa:1}, but at $\mathrm{[Fe/H]}=0.2$ for mixtures
[$\alpha$/Fe] = 0.0 ($p02p00$; solid lines) and [$\alpha$/Fe] = 0.4 
($p02p04$; dotted lines).}
\protect\label{fa:2}
\end{figure}

In Figs.~\ref{fa:1} and \ref{fa:2} we show in the
Hertzsprung-Russell diagram (HRD) the evolutionary tracks
for mixtures of identical [Fe/H] with solar and $\alpha$-enhanced metal ratios.
\footnote{The case $\mathrm{[Fe/H]} =
0.0$ has already been presented in \cite{weiss+07}.}  In the first
case ($\mathrm{[Fe/H]} = 
-0.5$) the hydrogen and helium abundances are very similar, such that
the effect of the enhanced total metallicity for the $\alpha$-enhanced
mixture shows clearly up in lower effective
temperatures and MS luminosities,  and thus longer
lifetimes. For example, the $1\,\msun$ has a turn-off age of 5.8~Gyr
(solar) and 7.4~Gyr ($\alpha$-enhanced).

On the other hand, the comparison at $\mathrm{[Fe/H]} = 0.2$ shows
comparatively small differences in the HRD; also the lifetimes of the
$1\,\msun$ star (and of the others) is quite similar (10.1 and
10.7~Gyr). The reason for this is that although the metallicity in the
$\alpha$-enhanced case is 0.048, compared to 0.026 of the solar-scaled case, the
hydrogen content is much higher in the latter (0.708 compared to
0.642), such that the lack of opacity due to the metals is compensated
for by the increased hydrogen opacity. This emphasizes the fact
already mentioned in \cite{wpm:95} that at high metallicities the
detailed assumptions about the galactochemical history of the helium
content becomes crucial. 

Thus, the choice of helium enrichment is becoming an issue for such
metal-rich systems. We decided to assume a moderate value for
the helium enrichment factor $\triangle Y/\triangle Z$, 
starting from a primordial helium content
of approximately 0.240. With this, $\triangle Y/\triangle Z$ for the
solar-like mixture $(X,Z)=(0.718,0.016)$ is 1.62 ($p00p00$ in Table 1), 
and for the
corresponding $\alpha$-enhanced one 1.53 ($p00p04$). For the other mixtures it
lies between 1 and 2, with the lowest value being 0.9 for the
$\alpha$-rich subsolar mixture. Given the uncertainty in the solar
model abundances and that of the galactic value for $\triangle
Y/\triangle Z$, and the question whether this value is the same for
solar-scaled and $\alpha$-enriched material at identical
$\mathrm{[Fe/H]}$, this choice is as good as any other.

\begin{figure} 
\includegraphics[scale=0.4]{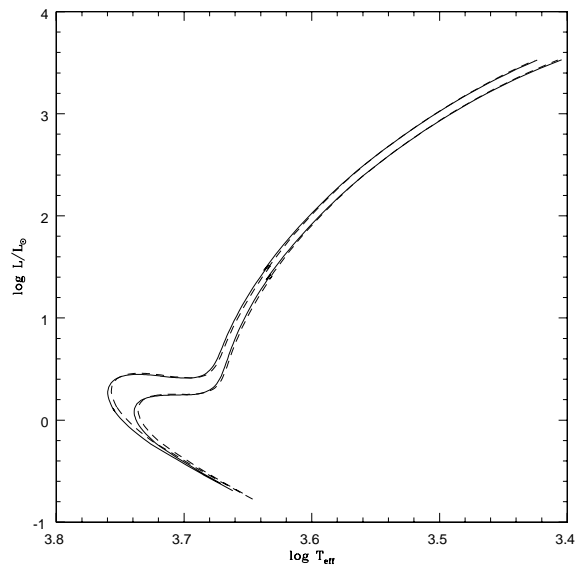}
\caption{Isochrones of 8 and 14 Gyr for models with $Z=0.032$ and
$\alpha$-enrichment, either variable (dashed lines) or constant 
(solid lines).}
\protect\label{fa:4}
\end{figure}

Another point of uncertainty is that of the actual
$\alpha$-enrichment factors. As discussed in \cite{weiss+07} the
difference between a constant or variable enhancement might lead to
crucial lifetime changes for stars of given mass. This is due to
systematic differences in the opacities at core temperatures. However,
it was also stressed that this refers to the evolution at fixed mass,
but not necessarily to isochrones at fixed age. We therefore show in
Fig.~\ref{fa:4} isochrones of 8 and 14~Gyr for models with either a
variable or constant $\alpha$-element enhancement. The differences are
not very large; e.g. the turn-off (TO) at 14~Gyr is hotter by less than
40~K, and fainter by less than 5 per cent for the variable enhancement
factors. However, the TO mass is $0.05\,\msun$ lower. We conclude that
for population synthesis aspects the influence of the rather
high sensitivity of the opacities with respect to composition
details is very modest and no source of concern.

\begin{figure} 
\includegraphics[angle=270,scale=0.32]{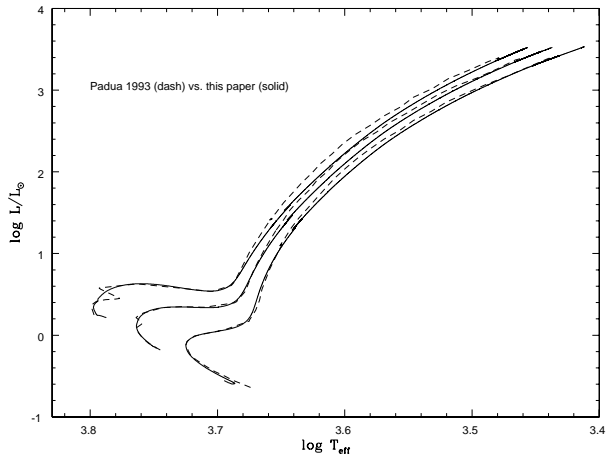}
\caption{Comparison between Bressan et al. (1993, dashed lines) and
our models (solid) for a
standard mixture of $X=0.70$, $Y=0.28$, and $Z=0.02$ for three stellar
masses (0.8, 1.0, and 1.2~$\msun$).}
\protect\label{fa:3}
\end{figure}

Finally, for comparison, we show in Fig.~\ref{fa:3} the tracks for a 
standard solar mixture of $(X,Z)=(0.70,0.02)$ for both the Padova library
(Bressan et al.\ 1993) and our own calculations. We compare only
low-mass stars, as the higher masses are affected even more by the
different treatment of overshooting. Even then, the effect is visible
in the structure around the turn-off, as our models all have only
transient convective cores during the earliest few million years, while
the Padova models for $M=1.0$ and $1.2\,\msun$ have persistent
convective cores. Otherwise, the tracks agree very well, given the
fact that almost 15 years lie between the two calculations. Only the
red giant branch (RGB) temperatures differ appreciably, which can be explained by the
fact that \citet{bressan+93} did not have access to the molecular
opacities by \cite{af:94}. Also the core hydrogen burning time of these 
models agree within 1.5 per cent.

\section{Stellar Spectral Energy Distributions}\label{stelsed}

The base of our stellar ingredient is the synthetic library by 
C05,
\footnote{\tt http://www.mpa-garching.mpg.de/PUBLICATIONS/{\par}DATA/SYNTHSTELLIB/synthetic\_stellar\_spectra.html}
which is
focused on the accurate reproduction of late-type stars. 
This library is based on the models atmospheres by \citet[ hereafter ATLAS9]{atlasodfnew},
and on a refined atomic and molecular line list calibrated through
several stellar spectroscopic studies \citep[see e.g.][]{barbuy+03}.
\footnote{Note on nomenclature: 
by {\it model atmosphere} we mean the run
of temperature and pressures as a function of optical depth which describes the 
photosphere of a
star. It distinguishes from {\it flux distribution} and 
{\it synthetic spectrum} which are the emergent fluxes calculated given a model 
atmosphere, whereas in general {\it flux distribution} corresponds to low-resolution
spectrum, and {\it synthetic spectrum} to high-resolution.}
We refer the reader to C05 for details on the high-resolution spectrum calculations. 

Some changes were made to the C05 library in order to better suit stellar
population studies: the spectra were smoothed with a gaussian
filter and resampled, so that we obtain spectra with a constant resolution of 
FWHM = 1~$\angstrom$ and a sampling of 0.2~$\angstrom$ pix$^{-1}$; 
additional models were computed to cover very cool 
giants, and smooth flux corrections were applied in order to 
improve the spectrophotometric predictions. 

The extension to cool giants and the flux corrections are described in detail below.

\subsection{Extension to cool giants}\label{c05ext}

C05 library covers from 3500 to 7000 K in stellar effective temperature. The lower
limit is set by the ATLAS9 model atmospheres, used as input in the 
spectral calculations. But it is well known that the RGB extends toward cooler stars, 
and these luminous giants have an important contribution to the integrated light
of old stellar populations. Therefore, 
an additional set of synthetic spectra was 
computed employing the \citet{plez92} MARCS models, for giants with 
temperatures down to 2800K, and adopting the same atomic and molecular 
line list and line formation code by C05. 

These MARCS models atmospheres do not account for $\alpha$-enhanced mixtures in
the opacities, nevertheless the $\alpha$-enhanced atomic and molecular abundances 
were properly
included in the calculation of the high-resolution spectra. We 
performed a test with a \Teff = 3500 K giant star from the ATLAS9 model 
atmospheres grid and we verified that this inconsistency 
has a negligible effect in the spectral 
line profiles, as long as the difference between the opacities of the model 
atmosphere and of the synthetic spectra is smaller then 
$\sim$0.4 dex. We also verified that the transition between ATLAS9-based spectra and 
MARCS-based spectra is smooth.

\begin{figure}
\includegraphics[width=\columnwidth]{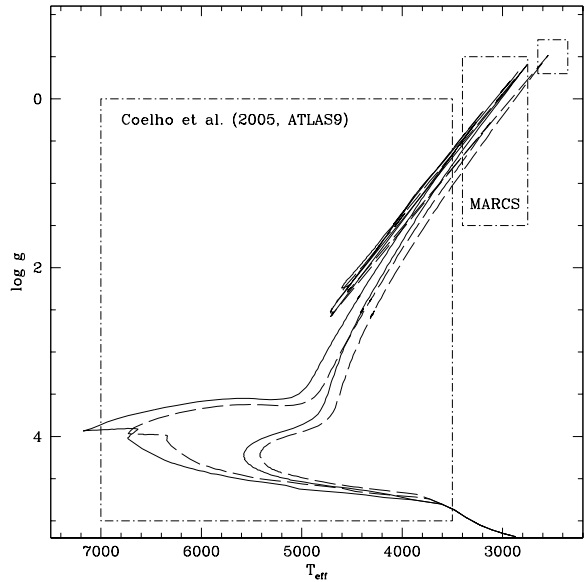}
\caption{Isochrones of 3 and 14 Gyr at fixed [Fe/H] = 0 are shown in {\Teff} 
versus {\it log g} parameter space. Mixtures [$\alpha$/Fe] = 0 and 0.4 are indicated 
as solid and dashed lines respectively. The large dot-dashed area shows the 
coverage 
of the original stellar library by C05 (based on ATLAS9 models),
and the two smaller dot-dashed areas show the 
extension computed in the present work (based on MARCS models). }
\label{fig_coverage}
\end{figure}

\begin{figure}
\includegraphics[width=\columnwidth]{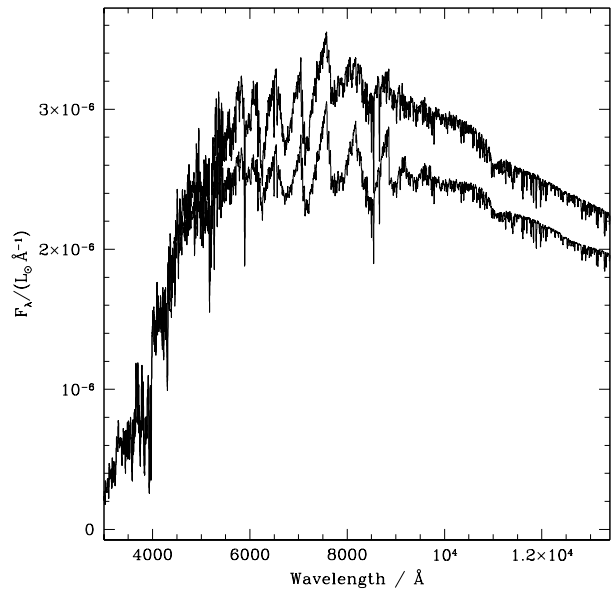}
\caption{ Spectral energy distribution of a 9 Gyr old SSP computed 
with and without the extension to cooler giants modelled for the 
present work (thin and thick lines, respectively).}
\label{fig_cool_giants}
\end{figure}

The coverage of the new stellar library is illustrated in  
Fig. \ref{fig_coverage} in the {\Teff} versus {\it log g} plane.
The effect of the inclusion of the very cool giants in the integrated
spectra of a 9 Gyr old SSP model is shown in Fig. \ref{fig_cool_giants}.

\subsection{Flux corrections}\label{fluxcalib}

Another update made to the original synthetic library concerns its 
spectrophotometry. A 
synthetic library that is intended to reproduce high-resolution line 
profiles with accuracy is not a library that also predicts 
good spectrophotometry. That happens because when computing a synthetic 
spectrum, one has to choose to include or not
the so-called 
`predicted lines' \citep[ hereafter PLs,][]{kurucz92}. The PLs are 
absorption lines for which one
or both energy levels have to be predicted from quantum mechanics
calculations.
They are an essential contribution
to the total line blanketing 
in model atmospheres and to spectrophotometric modelling. But as the 
quantum mechanics predictions are accurate to
only a few percent, wavelengths and computed intensities for these 
lines may be largely uncertain, and the PLs
may not correspond in position and intensity to the observable 
counterparts \citep{bell+94,castelli+kurucz04}. Therefore they are usually
not included in high-resolution spectral computations focused on 
spectroscopic use, which is the case
of the C05 library.

\begin{figure}
\includegraphics[width=\columnwidth]{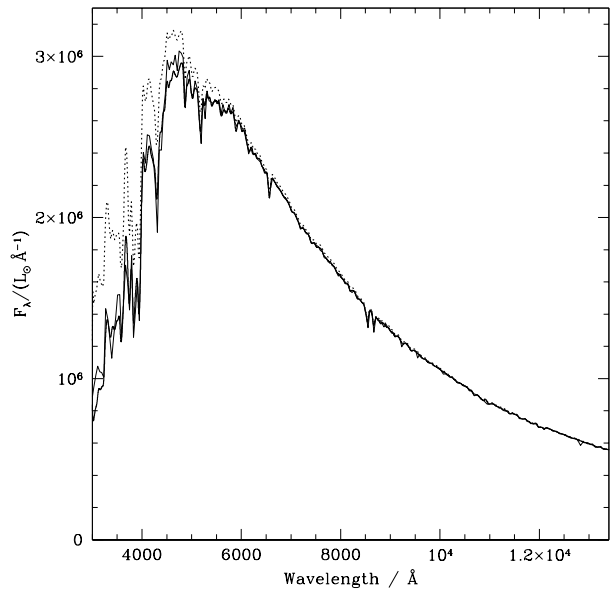}
\caption{ Spectral energy
distribution of a star with T$_{\rm eff}$ = 5750K, 
log g = 4.5, [Fe/H] = 0.0 and [$\alpha$/Fe] = 0.0.
The thin solid line shows the flux distribution as 
calculated by \citet{atlasodfnew}. The dotted line shows
the corresponding spectrum in the original C05 library
(the high-resolution spectra were smoothed to the resolution
of the Castelli \& Kurucz 2003 flux distributions).
 The thick solid line shows the spectrum adopted here, derived
from the first two spectra as described in the text.
}
\label{fig_PLcor}
\end{figure}

As a consequence, the colour predictions by C05
are inaccurate, in particular in the blue part of the spectrum. The grid 
by \citet{atlasodfnew} provides flux distributions 
predictions that include all the blanketing (i.e. the PLs). 
In order 
to correct the spectra by C05, each star in this 
library was compared to its counterpart flux distribution by 
\citet{atlasodfnew}. Both distributions were smoothed until no spectral 
feature remained. The ratio of the two curves describes how the C05 
synthetic star deviates from a model that includes all blanketing. This 
{\it response} was multiplied to the original spectrum given by C05, star by star. 
Therefore, the final spectrum kept the high-resolution features of the original 
C05 library, but presents a flux distribution which is closer to that 
predicted when including all blanketing. An original and corrected spectrum 
for a Sun-like star is shown in Fig. \ref{fig_PLcor}, in order to illustrate 
the effect of this correction.

\section{Population synthesis models}\label{popsyn}

The set of evolutionary tracks described in Section \ref{stelevol}
do not follow the evolution of the stars to the last
evolutionary phase for all masses. Therefore, 
the post-RGB evolution of the stars up to the
end of the early AGB were extended by using appropriate 
tracks by \citet[][ hereafter BaSTI]{pietrinferni+06}. (We are grateful to
S. Cassisi for providing us with a set of tracks
tailored to our needs.)
\footnote{Available at http://www.te.astro.it/BASTI/index.php}. 
As the effective temperatures of both
track sources to not agree completely along the giant branches, a
smooth switching from our to the BaSTI tracks over one brightness
magnitude was done. The temperature differences to be bridged were
typically around 100~K. 
The calculations by \citet{marigo+07} were additionally used to describe 
the thermally pulsing phase of the AGB (TP-AGB).
The BaSTI and \citet{marigo+07} tracks are not available for the exact mixtures
in Table 1, and we used instead the tracks with the closest value of Z.

The tracks were resampled to
identify the same equivalent physical stages required by the isochrone
synthesis technique as implemented by Charlot \& Bruzual (in preparation)(hereafter CB07).
The resampled tracks include 275 steps, distributed as follows: 55 along the MS,
25 in the sub-giant branch (SGB), 100 in the RGB, 
55 in the horizontal branch (HB), 25 along the AGB, and 15 in the
TP-AGB. 

By linear interpolation in the $(log~L,~log~T_{eff})$ plane we built isochrones
describing the locus in the HRD occupied by stars of each of the six
mixtures in Table 1 at ages from 3 to 14 Gyr in 1-Gyr steps.
The number of stars at each of the 275 evolutionary stages included in each
isochrone is computed from the IMF.
The spectrum of the star at each position was obtained by bilinear interpolation
in $(log~T_{eff},~log~g)$ in the corresponding metallicity plane in the C05 synthetic
stellar library, and then scaled according to the star surface area.
The integrated spectrum of the stellar population follows by adding the
stellar spectrum weighted by the number of stars at each position in the HRD.
Our final SSP models provide the integrated spectrum of the population at 
11 time-steps, in the wavelength range from 3000{\angstrom} to 1.34$\mu m$ with a 
constant resolution of FWHM=1{\angstrom} (at 0.2\angstrom pix$^{-1}$).

\begin{figure}
\begin{minipage}{\columnwidth}
\center
\includegraphics[width=\columnwidth]{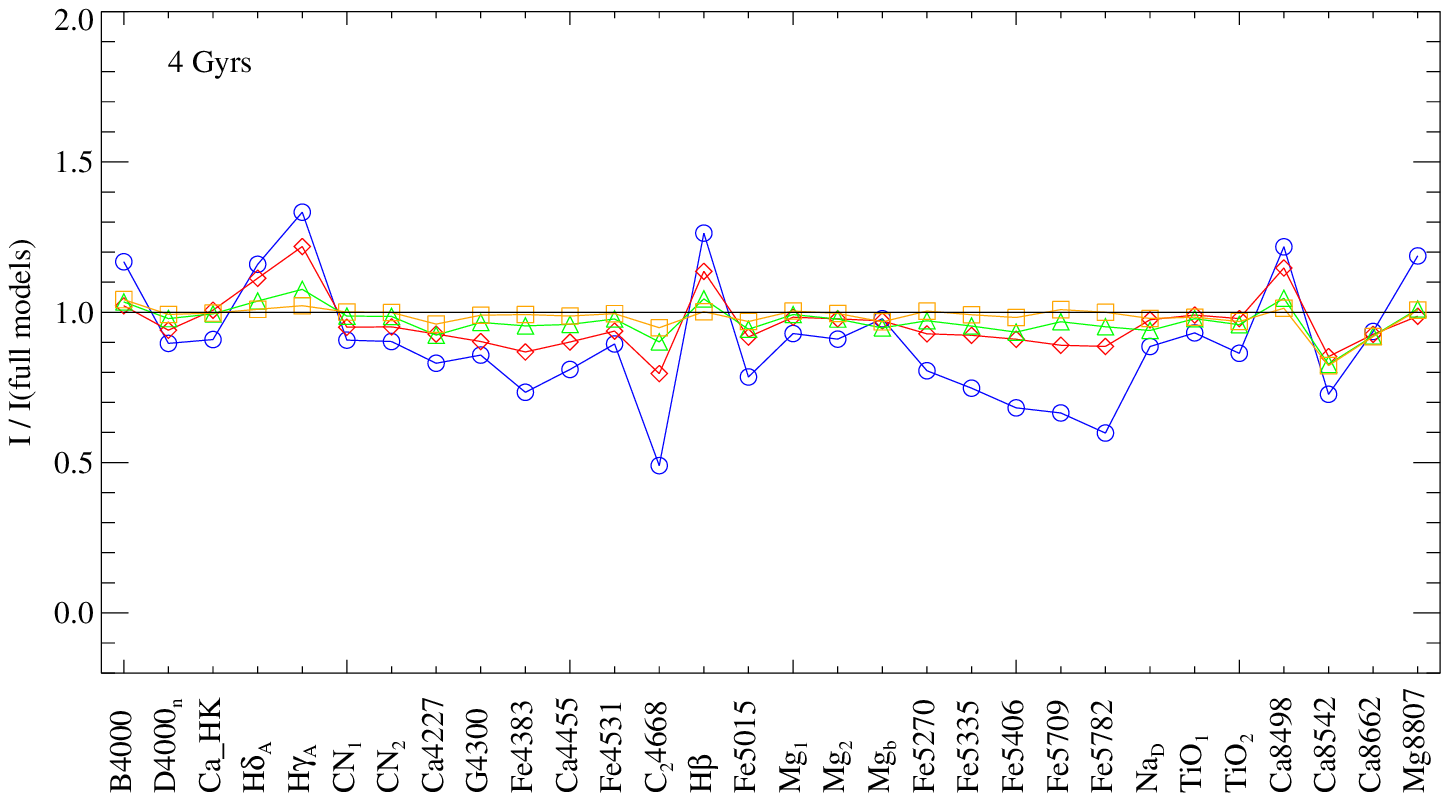}
\includegraphics[width=\columnwidth]{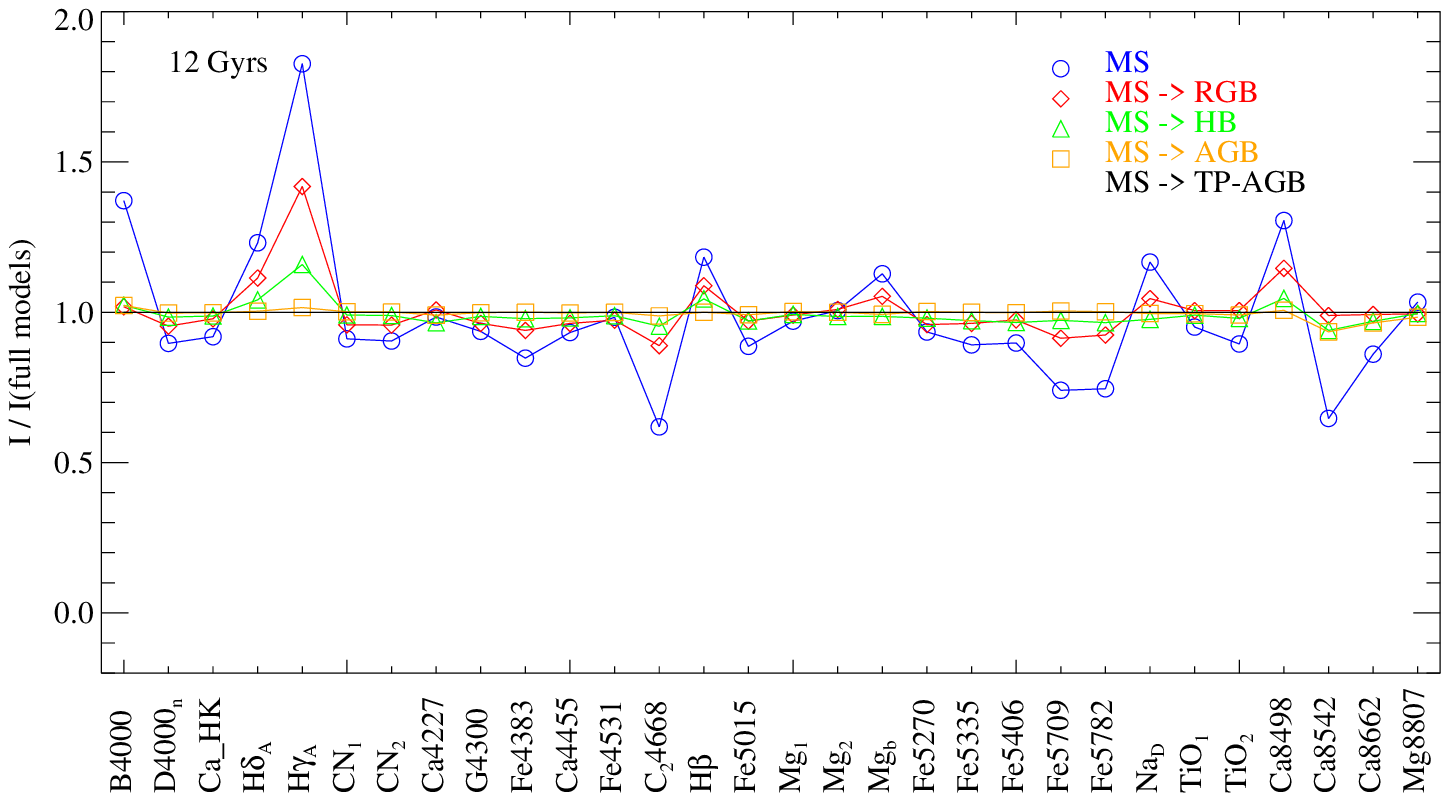}
\caption{ Lick-index strengths of model SSPs including different sets of stellar evolutionary 
phases (as indicated), in units of the index strengths of a model SSP including all phases (from MS to 
TP-AGB). The spectral indices are ordered in function of increasing central wavelength. The top and 
bottom panels show the results for a 4-Gyr and a 12-Gyr old populations, respectively. All models
have solar abundances ([Fe/H]= [$\alpha$/Fe] = 0).}
\label{fig_phases_effect}
\end{minipage}
\end{figure}

We provide two versions of the models. The {\it consistent} models include only 
the evolutionary phases MS, SGB and RGB. In this case, the assumptions for the chemical 
mixtures are fully consistent among the ingredientes, and these are the preferred
models to evaluate the differential effect of the $\alpha$ enhancement
on spectroscopic and photometric observables. 
The {\it full} models include the evolutionary phases MS, SGB, RGB, HB, AGB and TP-AGB,
and are more suitable to the direct comparison to observations. 
The effect of different evolutionary phases on the spectral indices is illustrated in Fig. 
\ref{fig_phases_effect}.

\begin{figure}
\begin{minipage}{\columnwidth}
\center
\includegraphics[width=\columnwidth]{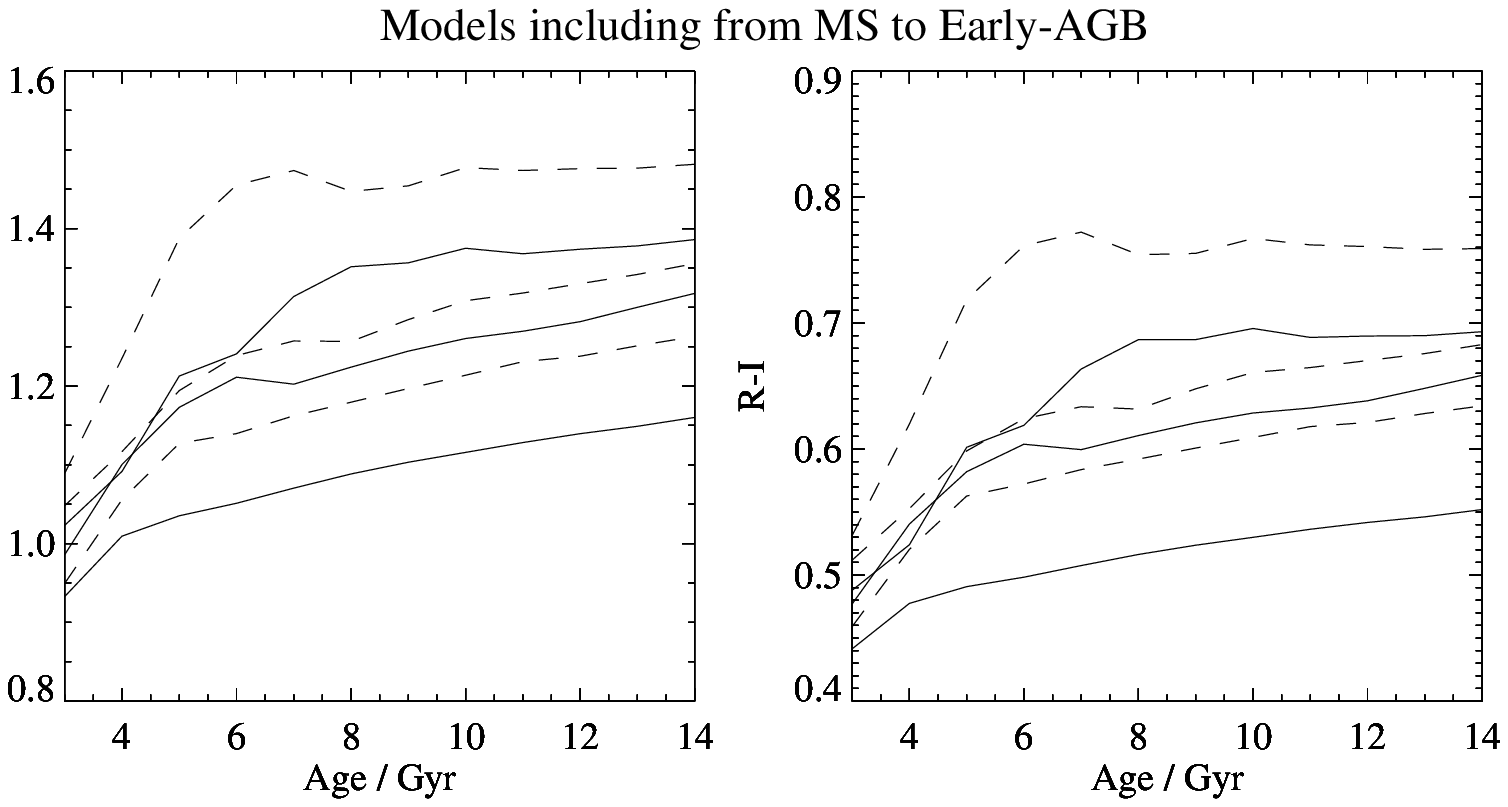}
\includegraphics[width=\columnwidth]{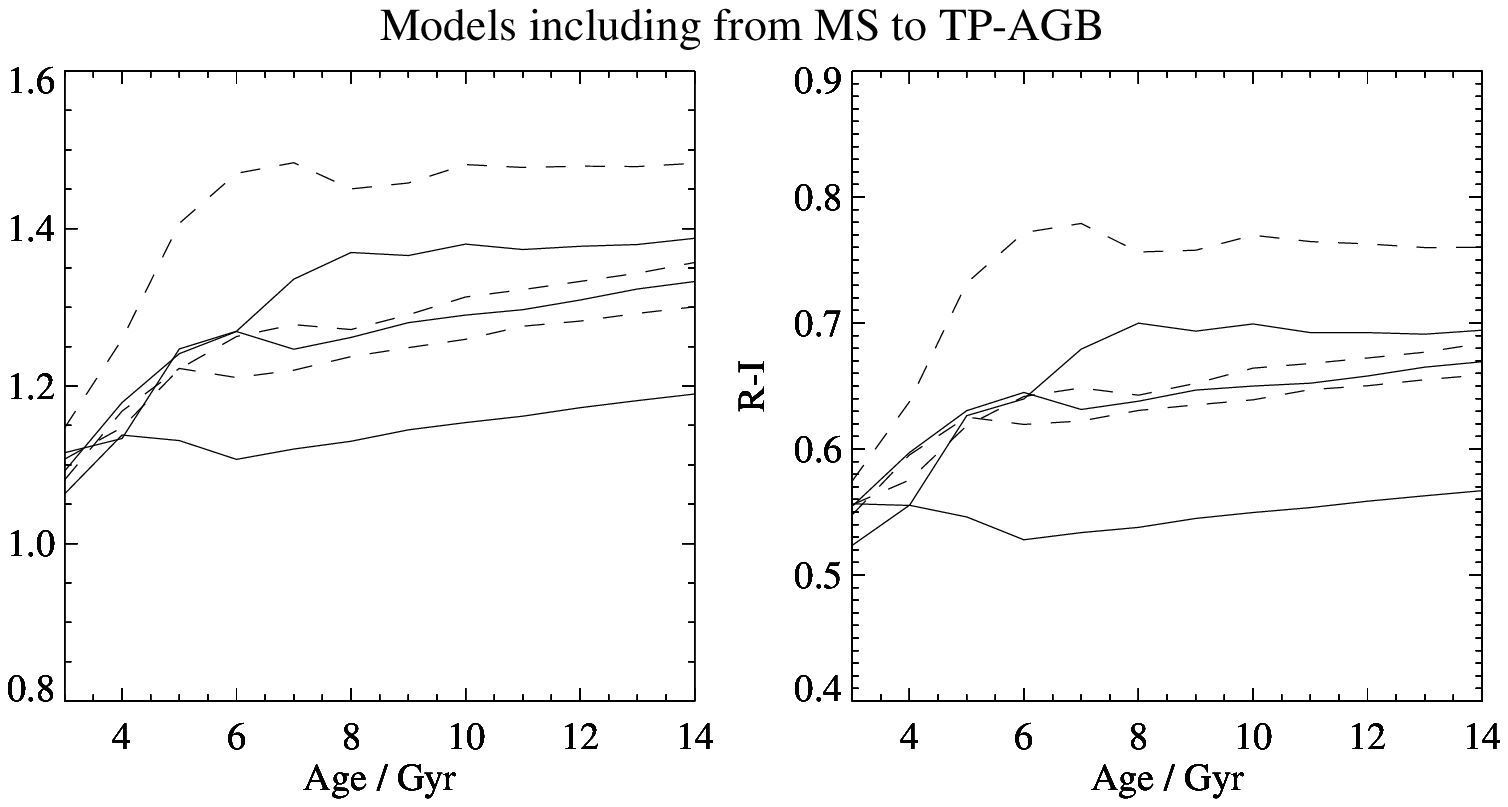}
\caption{Colour evolution predicted by our SSP models, with and
without the inclusion of the TP-AGB phase. The solid lines
show the evolution of the three solar-scaled mixtures, and the dashed
lines show the $\alpha$-enhanced mixtures. In each case the three iron 
abundances
are represented ([Fe/H] = $-0.5$, 0.0 and 0.2), and higher iron abundances
correspond to redder colours.}
\label{fig_tpagb_effect}
\end{minipage}
\end{figure}

We noted that the addition of the TP-AGB phase has a small effect on the I-band flux,
affecting therefore some visual colours, as can be seen in Fig. \ref{fig_tpagb_effect}.
This effect is mostly seen in the younger ages and solar-scaled mixtures (the
$\alpha$-enhanced populations are in comparison more populated with cool stars, and the addition
of a very cool TP-AGB spectrum is less significant).
We consider these results as preliminary indications only, for the
following reasons. The stellar spectra that represent the TP-AGB phase
are selected among the MARCS-based spectra (cf. Section \ref{c05ext}), which 
are not appropriate models to represent TP-AGB stars (that are known
to be either carbon or oxygen enriched).
Secondly, the effective temperatures of AGB stars are
uncertain due to several uncertainties in the stellar evolution models, most
notably convection in the highly super-adiabatic outer envelope, 
structure and radiation transport in the atmosphere, and finally
the composition, influenced by dredge-up and hot bottom burning,
itself. Thirdly, the mass loss, depending itself on these effects,
leads to circumstellar (dust) shells, which absorb shorter wavelength
light and reemits it in the far infrared. The properties of the
circumstellar shell again depend on chemistry and mass loss
history. Several studies, among them \citet{piovan+03} and \citet{groenewegen06}
have modelled that effect in the IR, and we expect therefore that, due to the
dense circumstellar dust shells, the contribution of
AGB stars to the visual and red bands will be smaller than our models
indicate, but larger in the mid- to far-IR bands, where dust
re-emits the absorbed radiation.

\section{The performance of fully synthetic SEDs}\label{synsed}

The major concern of using a synthetic stellar library as ingredient for 
SSP models regards the accuracy of the model stars when compared to 
observed ones. For this reason, the large majority of SSP models to date employ empirical stellar
libraries. In order
to assess the impact of using a synthetic stellar library 
on SSP models,
we performed some comparisons between fully theoretical 
SSP models and
models from literature that use empirical libraries
(a detailed comparison between 
theoretical and empirical libraries in a star by star basis is given in
Martins \& Coelho 2007).
It is worth noting that
empirical libraries are not error-free, being affected by 
uncertainties on the stellar atmospheric parameters, 
spectrophotometric calibration issues, poorly constrained
chemical pattern, telluric contamination, and other effects 
difficult to control. We aim here at highlighting 
the major differences, 
rather than performing a definitive evaluation of the quality of the 
synthetic library.

\subsection{Comparisons of spectral energy distributions}

{\b We computed SSP models (hereafter SYN models) with the synthetic spectral
library described in Section \ref{stelsed} and compared them with models by BC03,
which employ the STELIB \citep{stelib} library, and models by CB07 built with
the Indo-US library \citep{indous}.
In order to isolate the influence of the stellar spectral library alone, 
all models considered in this section employ the
isochrones by \citet{girardi+02} and a \citet{salpeter} IMF.}

\begin{figure}
\begin{minipage}{\columnwidth}
\center
\includegraphics[width=\columnwidth]{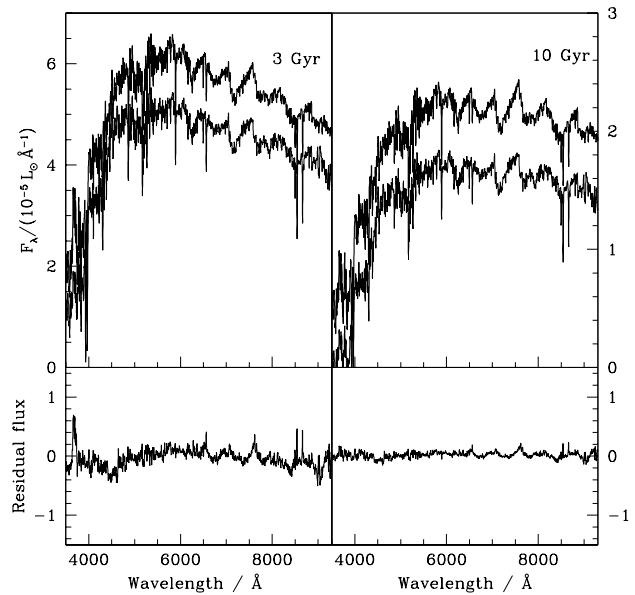}
\caption{{\it Top:}  Comparison of SSP models computed using the library
of synthetic stellar spectra described in Section \ref{stelsed} (SYN models; thick lines)
and BC03 models (thin lines), at the ages of 3 Gyr ({\it left}) and 10 Gyr ({\it 
right}). The BC03 spectra were arbitrarily shifted to facilitate visualization. 
{\it Bottom:} Residuals between the two models. All models were computed with the
solar isochrone by \citet{girardi+00} and a Salpeter IMF.}
\label{fig_compac1}
\end{minipage}
\end{figure}

\begin{figure}
\begin{minipage}{\columnwidth}
\center
\includegraphics[width=\columnwidth]{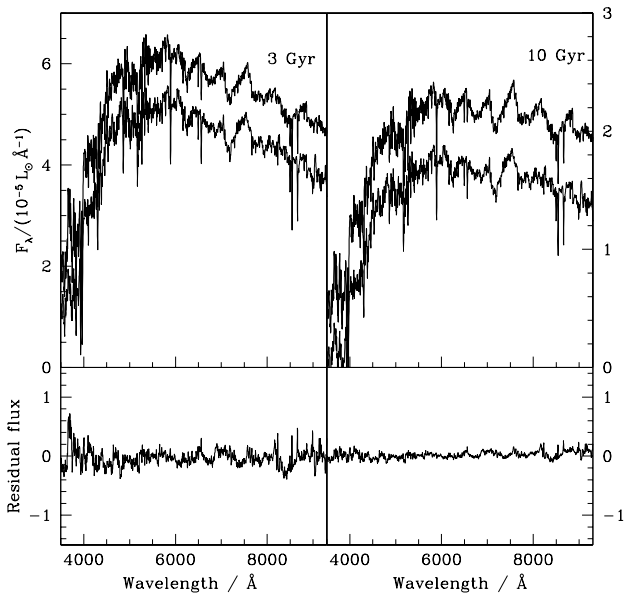}
\caption{Same as in Figure \ref{fig_compac1}, but showing models by CB07
using the Indo-US library as thin solid lines.}
\label{fig_compac2}
\end{minipage}
\end{figure}

The  model SEDs distributions are shown in 
Figs \ref{fig_compac1} and \ref{fig_compac2}, comparing SYN models to the ones by
BC03 and CB07, respectively,  at the ages of 3 and 10 Gyr. In each case, the
top panel shows the compared spectra and the bottom panel the residuals.  Residuals
are on average smaller 
for the old age population. In the case of the 3-Gyr
populations, there is a residual peak centered around 3700 $\angstrom$ 
that is not present in the older age. It comes from the fact that 
C05 library includes only the first 10 Balmer lines in its spectrum, and as 
a consequence, the hotter stars in that library present a misfeature 
around 3700 $\angstrom$ due to missing higher
order Balmer lines.
There are also peaks at longer wavelengths (from 6500$\angstrom$ on) that are
related to small mismatches in the intensity of the TiO bands. From 
the theoretical point of view, the spectrum of this complex molecule 
is difficult to model, and C05 did a careful empirical calibration 
of the intensity of the TiO bands against cool giants. From the point 
of view of the empirical library, this is a region dominated by cool
giants, whose atmospheric parameters are largely uncertain. Also, TiO 
intensity is very sensitive to Ti and O abundances, and the observed 
stars from the solar neighbourhood may show variations on [Ti/Fe] 
and [O/Fe] \citep[e.g.][]{reddy+03}, while the synthetic library has fixed values, 
by construction.

\subsection{Broad-band colours}

\begin{figure}
\includegraphics[width=\columnwidth]{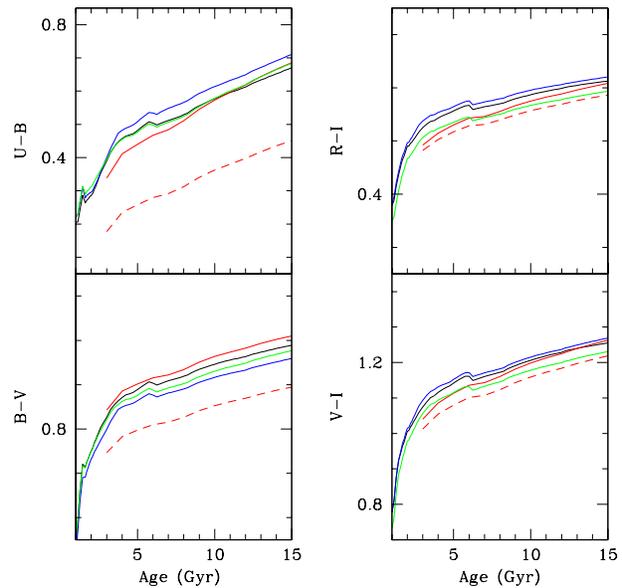}
\caption{Colour evolution of SSPs adopting different stellar libraries. 
Models using empirical libraries are shown in black (STELIB) and blue (Indo-US). 
In 
green is shown the predictions by the BaSeL low-resolution synthetic 
library. Red solid lines are SYN models of the present work. The dashed red lines
indicate models computed with the original C05 library, i.e., prior to the
flux corrections explained in Section 3.2.
}
\label{fig_johnson}
\end{figure}

To asses the spectrophotometric properties of the fully synthetic 
models, we computed broad-band colours and compared them to the predictions 
of BC03 and CB07 models. 
The results are presented in Fig. \ref{fig_johnson}, showing 
the age evolution of the $\rm (UBV)_{Johnson}-(RI)_{Cousins}$ colours given
by SYN models (solid red lines), BC03 models (with STELIB library; black lines), 
CB07 models (with Indo-US library; blue lines), and an additional 
set of BC03 models using the BaSeL theoretical library \citep[][green lines]{basel1, 
basel2, basel3}.  The dotted red lines indicate the colours  obtained using the 
original C05 library, prior to the flux corrections described in Section \ref{fluxcalib}. 
 Fig. \ref{fig_johnson} shows that the flux corrections improve drastically the 
synthetic colours. The SYN colours seem to be in general 
marginally bluer than the average prediction given by empirical libraries
(except in the B$-$V colour, that shows the opposite behaviour), but the 
differences are small. We conclude that the modified version of C05
library is fairly accurate in predicting broad-band colours, 
showing no large differences compared to models based on
empirical libraries. 

\subsection{Classical spectral indices}\label{synsedidx}

\begin{figure*}
\begin{minipage}{\textwidth}
\center
\includegraphics[width=14cm]{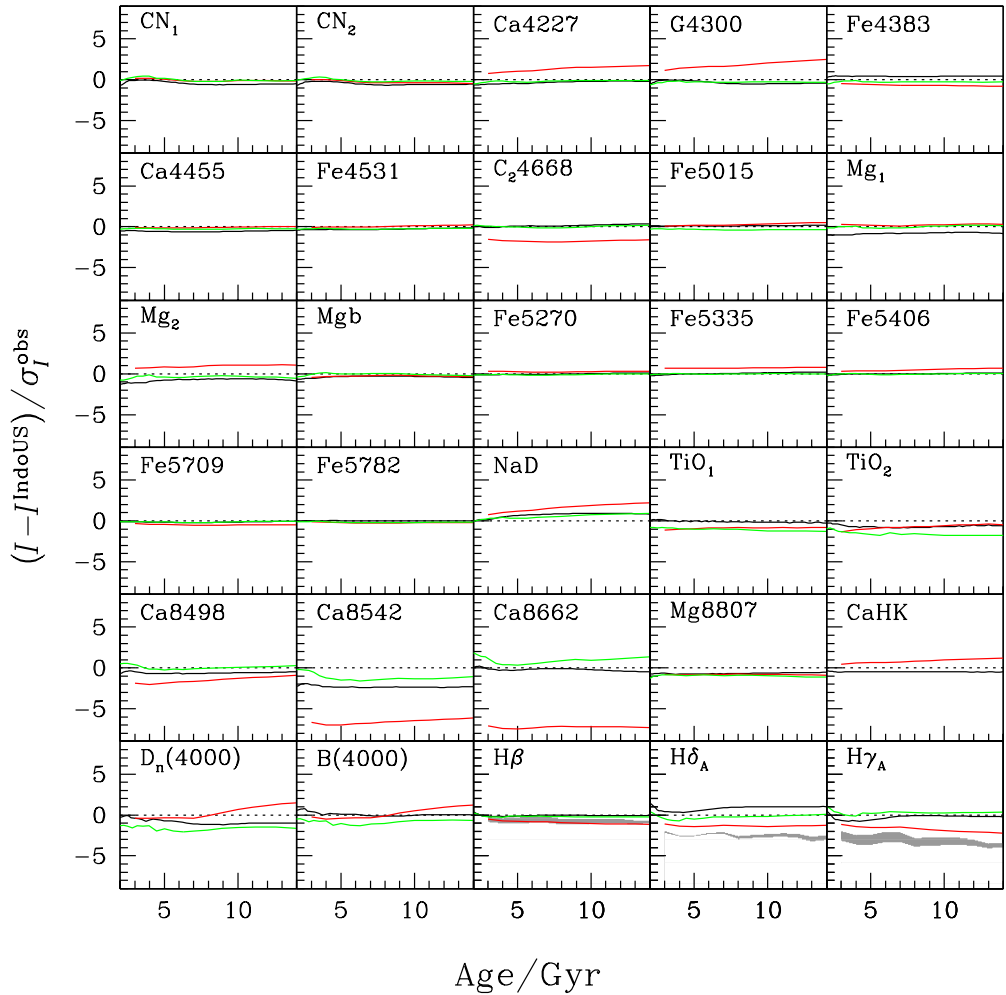}
\caption{Evolution of spectral indices, for SSP models computed using different stellar 
libraries. BC03 models are shown in black, SYN models in red and Vazdekis 
models 
(built with either MILES or Cenarro et al. 2002 libraries) in green.  For each index,
the {\it y}-axis shows 
the difference between a given SSP model and the CB07 Indo-US model, in units of the
typical observation error (Table \ref{tab_errors}).
In the panels that show Balmer line indices, the shaded region shows the 
values predicted by other two high-resolution theoretical libraries 
\citep{martins+05,munari+05}.}
\label{fig_allidx}
\end{minipage}
\end{figure*}

 To assess the accuracy of the synthetic library in
predicting some of the widely used spectral indices, 
we measured a group of 30 indices on SYN models. 
These indices were compared to the predictions by BC03, 
CB07 (with Indo-US library), 
\cite{vazdekis+03}\footnote{These models are based on \cite{cenarro+02} 
stellar fitting functions and cover 
the near-IR wavelength range.} and
Vazdekis et al. (in preparation)\footnote{Based on the recent MILES library by 
\citet{MILES,cenarro+07}, covering the optical wavelength
range.} models.
The selected indices encompass the classical 
Lick/IDS indices as defined by \citet{trager+98}, the higher order 
Balmer lines by \citet{worthey_ottaviani97}, the near-IR indices by \citet{diaz+89}, 
two definitions of the D4000 break \citep{bruzual+83,balogh+99} and the blue Ca H\&K index by
\citet{brodie_hanes86}.
All models were smoothed 
to a common resolution of 70~km s$^{-1}$ prior to the measurement 
of the indices.
The results are shown in Fig. \ref{fig_allidx}, which shows the 
differences of model indices  relative to Indo-US models by CB07, in units of the
observational 
errors. The observational errors are taken from red galaxies in SDSS-DR4 
with S/N $\ge$ 40  per pixel, and are listed in Table \ref{tab_errors}.
For most of the indices the agreement is very good. The larger discrepancies 
are discussed below.

\begin{table}
\begin{minipage}{\columnwidth}
\caption{Observational errors adopted in Figure 
\ref{fig_allidx}.}
\label{tab_errors}
\begin{center}
\begin{tabular}{lc}
\hline
Index & Observational error \\
\hline
B4000		& 0.062\\
D$_n$(4000)	& 0.054\\
Ca H\&K		& 0.040\\
H$\delta _A$& 0.942\\
H$\gamma _A$& 0.908\\
CN$_1$      & 0.027\\
CN$_2$   	& 0.908\\
Ca4227	 	& 0.452\\
G4300 	 	& 0.809\\
Fe4383	 	& 1.063\\
Ca4455	 	& 0.506\\
Fe4531	 	& 0.816\\
C4668 	 	& 1.286\\
H$\beta$ 	& 0.524\\
Fe5015	 	& 1.275\\ 
Mg$_1$   	& 0.016\\
Mg$_2$   	& 0.021\\
Mg$_b$   	& 0.653\\
Fe5270	 	& 0.673\\
Fe5335	 	& 0.690\\
Fe5406	 	& 0.535\\
Fe5709	 	& 0.376\\
Fe5782	 	& 0.336\\
Na$_D$   	& 0.460\\
TiO$_1$  	& 0.012\\
TiO$_2$  	& 0.010\\
Ca8498   	& 0.162\\
Ca8542	 	& 0.156\\
Ca8662	 	& 0.179\\
Mg8807	 	& 0.129\\
\hline					   
\end{tabular}
\end{center}
\end{minipage}
\end{table}

{\it Balmer line indices:}
All the Balmer indices are underestimated in the SYN 
models. This effect was expected because it is known that 
Hydrogen lines computed in local thermodynamic equilibrium (LTE) match well the 
wings, but cannot reproduce the core of the lines. Better suited microturbulence 
velocities and/or a mixing length to pressure scale height ratio $\ell$/H$_{\rm p}$ 
improve somewhat the match 
\citep[e.g.][]{canuto+92,fuhrmann+93,vantveer+96,bernkop98,barklem+02},
and this is the reason why \citet{barbuy+03} recomputed the ATLAS9 model 
atmospheres with $\ell$/H$_{\rm p}$  = 0.5, but a single value is hardly sufficient
for all spectral types. 
Besides, the bottom of the hydrogen lines form in the chromosphere, not included 
in the model atmospheres. In the 
very cool stars these lines are weak and have essentially no wings, and as a consequence
the indices are dominated by a non-satisfactory line core modelling.
We built models based on two other recent synthetic libraries 
\citep[ shown as a shaded 
region in Fig. \ref{fig_allidx}]{martins+05, munari+05}, and we confirm that 
these stellar models 
show the same systematic behaviour. 

{\it Indices sensitive to C and N abundances:}
Among the indices sensitive to C and/or N abundances, G4300 and C$_2$4668
deviate more than 3 sigmas from the predictions based on empirical
libraries. We believe this happens partially because in C05 library,
the C and N abundances relative  relative to Fe were assumed to be solar through
all evolutionary phases.
But it is well known that the CNO-cycle lowers the C abundance and enhances
the N abundance in giants \citep[e.g.][]{iben67,charbonnel94}. The same effect
on the indices CN$_{1}$ and CN$_{2}$ would not be so clearly seen because in 
these indices the variations of C and N somewhat compensate each other. 
The 
calculation of a sub grid of the synthetic models accounting for C and N 
abundance variations is underway and will be presented in a separate paper 
\citep[with similar assumptions to the ones adopted in]
[ and additionally including consistent computations for the model atmospheres]
{barbuy+03}.
It is interesting to note that another index that deviates by similar amounts 
is Ca4227. Despite its designation, the \citet{TB95} tables show that the sensitivity of this 
index to C variations is of the same order as its sensitivity to Ca abundance 
variations.

{\it Near-IR Ca Indices:}
The near-IR indices show the biggest deviation 
among all the indices (Ca8542 and Ca8662 deviate beyond 5 sigmas in error units). When we 
compared the C05 calculations with very high resolution observed spectra of 
the Sun and Arcturus \citep[]{solarflux:kurucz84,arcturusflux:hinkle00}, we  
found that the synthetic indices were underestimated 
\citep[as expected from the lack of N-LTE line profile calculation, see e.g.][]{cayrel+96}, 
but the differences were typically 3 times smaller than the 
difference shown in Fig. \ref{fig_allidx}.
The larger difference found in the SSP models is possibly coming 
from a source different than only inaccuracies in the synthetic stars. 
We noted that \citet{bensby+05} found that the disk stars with 
[Fe/H] = 0.0 $\pm$ 0.1 tend to have [Ca/Fe] values in the range between 0.0 and 0.1,
so it is possible that models built with empirical libraries are contaminated
by stars with [Ca/Fe] $\neq$ 0.0, that would prevent a proper match
by SYN models\footnote{The solar models 
by BC03 include stars from STELIB with $-0.06$ $\leq$ [Fe/H] $\leq$ +0.07. 
CB07 models adopt the range $-0.10$ $\leq$ [Fe/H] $\leq$ +0.10.}.
This effect would not be so clearly seen
in the blue Ca indices because the response of the red Ca indices to [Ca/Fe] is considerably
stronger see (Appendix A in the Supplementary   
Material, available only in the online version of this paper). 
Additionally, this wavelength range is dominated by cool giants, whose stellar 
parameters in the empirical libraries are largely uncertain, and \citet{martins_coelho07}
show that these indices are also highly sensitive to flux-calibration issues.
We conclude that the large discrepancy is related both to uncertainties
in the synthetic calculations (due to N-LTE effects) and uncertainties in the empirical
libraries (abundance pattern, atmospheric parameters and flux calibration issues).

\section{Model predictions}\label{modpred}

In this section we present the new spectral models, computed with 
the ingredients described in Sections \ref{stelevol} and \ref{stelsed}, the synthesis code
presented in Section \ref{popsyn} and adopting IMF by \citet{chabrier03}.
The space parameter covered by the models is summarized below:

\begin{description}
\item $-$ wavelength range: from 3000~{\AA} to 1.34$\,\mu$m
\item $-$ [Fe/H]: $-$0.5, 0.0 and 0.2
\item $-$ [$\alpha$/Fe]: 0.0 and 0.4
\item $-$ Total metallicity Z: 0.005, 0.011, 0.017, 0.026, 0.032, 0.048 
\item $-$ Ages: 3 to 14 Gyr (1-Gyr step)
\end{description}

The original resolution of the models is FWHM = 1 $\angstrom$ 
with a sampling of 0.2 $\angstrom$ pix$^{-1}$, and we provide additionally 
models convolved to the wavelength dependent instrumental 
resolution of SDSS. 
Different resolutions and samplings can be obtained by appropriately
convolving the original models.

Some applications of the models are presented below.

\subsection{Classical Mg versus Fe plot}

\begin{figure*}
\begin{center}
\includegraphics[width=10cm]{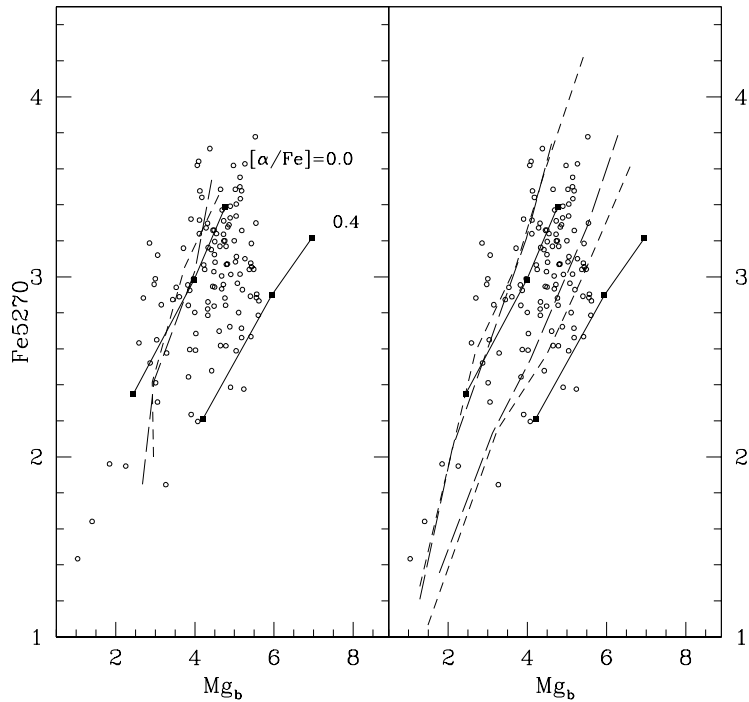}
\caption{SSP models at constant age of 12 Gyr are plotted with 
the galaxy data from \citet{trager+98}.
{\it Left-hand panel:} Solid lines 
show models by the present work, where the solid squares indicate 
[Fe/H] = $-$0.5, 0.0 and 0.2 (the most metal-rich have higher index values). 
Short dashed lines indicate BC03 models, and long dashed lines are CB07 Indo-US models. 
{\it Right-hand panel:} The solid line is the same as in the left-hand panel. 
Short dashed lines are models by \citet{TMB03}, also for [$\alpha$/Fe] 0.0 and 0.4. 
Long dashed lines indicate models by \citet{schiavon07}.}
\label{fig_classicalplot}
\end{center}
\end{figure*}

The best known result that illustrates the effects of $\alpha$ enhancement 
on spectral properties of galaxies 
are those of magnesium versus iron indices \citep[e.g.][]{wfg92}, where
solar-scaled models cannot reproduce the locus of high-mass spheroidal galaxies.
In Fig. \ref{fig_classicalplot} we present our models for 12-Gyr populations 
(solid lines) in the {\it Fe5270 versus Mg $b$} plane, with
galaxies from \citet{trager+98} sample, which span a large range in masses.

The indices were measured at the Lick/IDS resolution as tabulated in 
\citet{worthey_ottaviani97}.
The synthetic stellar library employed in our models is closer to a 
flux-calibrated 
system than to the Lick/IDS flux system. We translated our measurements
to the standard Lick/IDS adopting the zero-point corrections
tabulated in BC03 (which translates the flux-calibrated stellar 
library STELIB to the Lick/IDS system). 

In the left-hand panel of Fig. \ref{fig_classicalplot}, the short-
and long-dashed lines are the predictions given by BC03 and CB07 models, 
respectively. 
It can be seen that the present models successfully cover the gap left by previous 
spectral models, i.e. by producing models that fill the locus of the massive 
early-type systems.  

In the right-hand panel we show the models of the present work and the 
Lick/IDS model indices by \citet{schiavon07} (short dashed lines) and 
\citet[ long dashed lines]{TMB03}, so that we can see how our models compare 
with Lick/IDS models available in the literature. 
The behaviour of all $\alpha$-enhanced models shown in 
Fig. \ref{fig_classicalplot} is similar, 
in the sense that the $\alpha$-enhanced models have stronger Mg indices when 
compared to iron indices. In detail, our models tend to produce stronger Mg 
indices at the same adopted mixture of [$\alpha$/Fe] = 0.4.

Besides the different ingredients and methodologies, there is a particular
difference in our approach of accounting for the $\alpha$ enhancement, 
with respect to the ones available in literature, that is worth highlighting. 
Lick/IDS model indices \citep[like those of ][]{TMB03,schiavon07} use the response 
functions to include the different abundance patterns. 
When deriving the response functions \citep{TB95,houdashelt+02,korn+05}, 
the calculations of synthetic spectra
consider {\it individual} element
enhancements, whereas other work 
\citep[e.g.][]{barbuy+03,atlasodfnew,coelho+05,munari+05,phoenix05,MARCS05}, 
as well as the present models,
assume {\it all} $\alpha$ elements enhanced
together. It is not clear if the effect of enhancing
all $\alpha$ elements is equivalent to
a linear combination of the effects
due to individual abundance variations.
That is because in stars of spectral type G to M, the continuum forms mainly by free$-$free 
and bound$-$free transitions of H$-$. The amount of H$-$ present
depends on the electrons captured by Hydrogen atoms, coming
in good part from $\alpha$ elements that are important electron donors.
Therefore one of the main effects of $\alpha$ element enhancement is a lowering
of the continuum, and a clear example is the weakening of the Fe5270 and Fe5335 
indices in spectra with $\alpha$ element enhancements, 
at fixed iron abundances \citep[e.g.][]{barbuy+03}. 
In fact all the spectrum gets weaker from the lowering of the 
continuum, while atomic and molecular lines involving $\alpha$ elements
get stronger due to the higher abundance. In some
cases, like that of the Mg$_2$ index, the
stronger intensity of the MgI + MgH lines present in the Mg$_2$ region
overcome the lowering of the continuum,
and the index becomes stronger with $\alpha$ enhancement 
\citep[e.g.][]{barbuy94}. 
Therefore, it seems to us that an overall enhancement of 
the $\alpha$ element abundances
can have an impact on the continuum (and therefore in the indices) that
is different from combining the impact of varying the abundance of each
element individually, through the response functions.

\subsection{Indices insensitive to $\alpha$/Fe}

\begin{figure}
\includegraphics[width=\columnwidth]{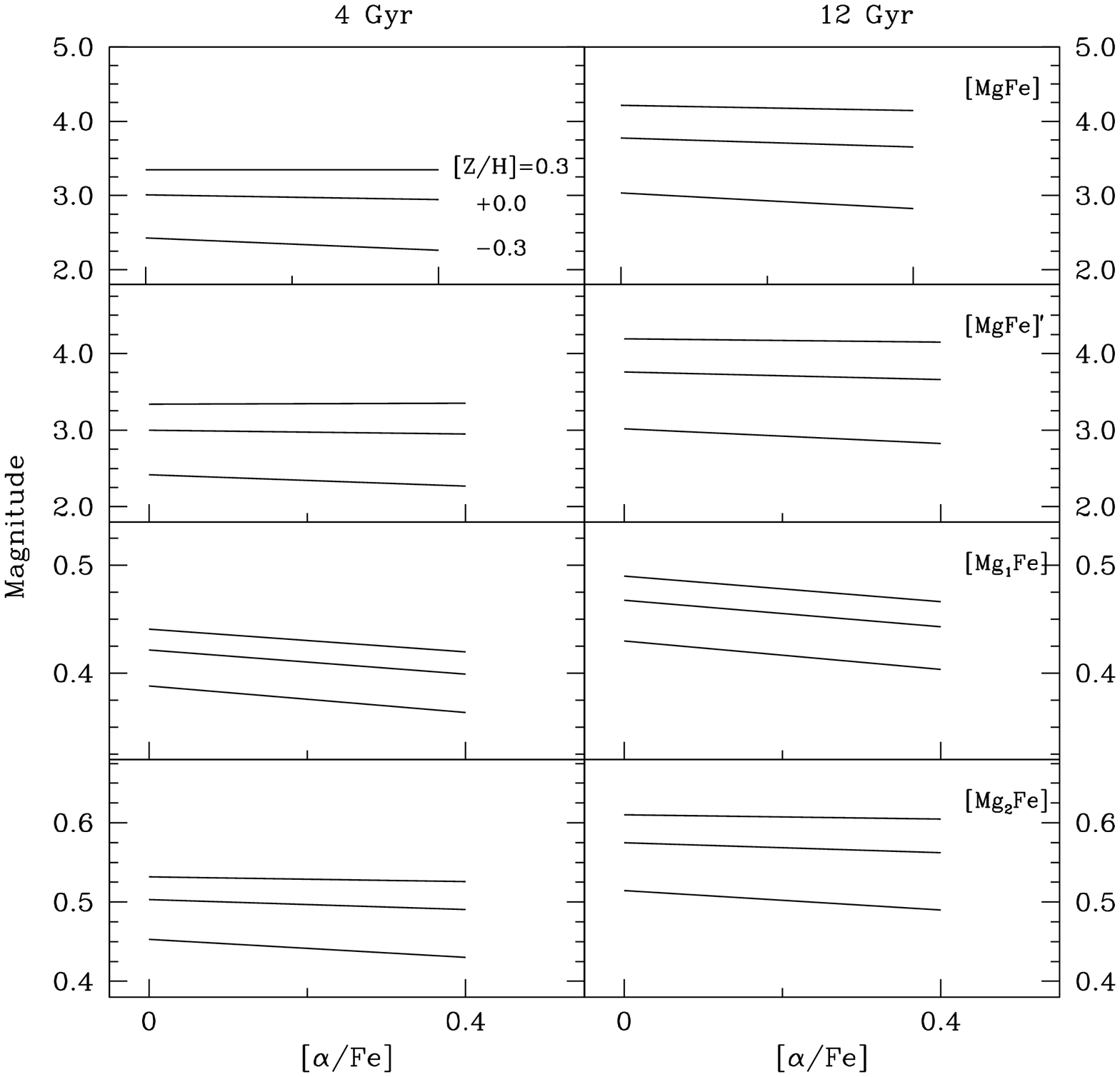}
\caption{The behaviour of four indices defined to be insensitive to 
[$\alpha$/Fe], 
given by \citet{gonzalez93}, \citet{TMB03} and BC03. The behaviour for two ages and 
three 
different values of [Z/H] were inspected.
The models for 4 Gyr are shown in the left-hand panel and models for 12 Gyr are shown
in the right-hand panel. Each row shows the behaviour 
of an index, as indicated in the top right hand corner of each panel. The three 
lines in 
each panel correspond to [Z/H] = $-$0.3, 0.0 and 0.3, from bottom to top.}
\label{fig_mgfeidxs}
\end{figure}

Indices defined to be insensitive to $\alpha$/Fe, providing a
good measurement of the total metallicity, have been used frequently
in the literature. 
Most of the SSP models available in literature 
adopt a {\it fixed Z} approach, as proposed by \citet{trager+00a} 
(the total metallicity Z is the same for the solar-scaled and for the $\alpha$-enhanced model). 
In order to keep the total metallicity constant, 
the iron abundance of the $\alpha$-enhanced models has to be lowered
({\it iron depletion}). 
The present models, being computed at fixed
[Fe/H], do not directly provide predictions at fixed Z. 
For the purpose of exploring some of the total metallicity indicators,
we interpolated a small set of our models to
estimate indices for three values of [Z/H] ($-$0.3, 0.0, +0.3)
and two ages (4 and 12 Gyr).
We studied four indices:
[MgFe] as given by \citet{gonzalez93}, 
[MgFe]' by \citet{TMB03} and 
[Mg$_1$Fe] and [Mg$_2$Fe] by BC03; and the results are shown   
in Fig. \ref{fig_mgfeidxs}. 
We find that the [Mg$_1$Fe] index gets smaller with increasing [$\alpha$/Fe], 
for any age or Z calculated. For the three remaining indices, [MgFe], [MgFe]' 
and [Mg$_2$Fe], we confirm that they are fairly insensitive to [$\alpha$/Fe]. 
There is a hint that at subsolar values of Z there is some dependence with
$\alpha$/Fe, but it is rather small. 
We see no appreciable difference among the behaviour of these three indices.

\subsection{Effect on colours}\label{modcol}

\begin{figure*}
\begin{center}
\includegraphics[width=11cm]{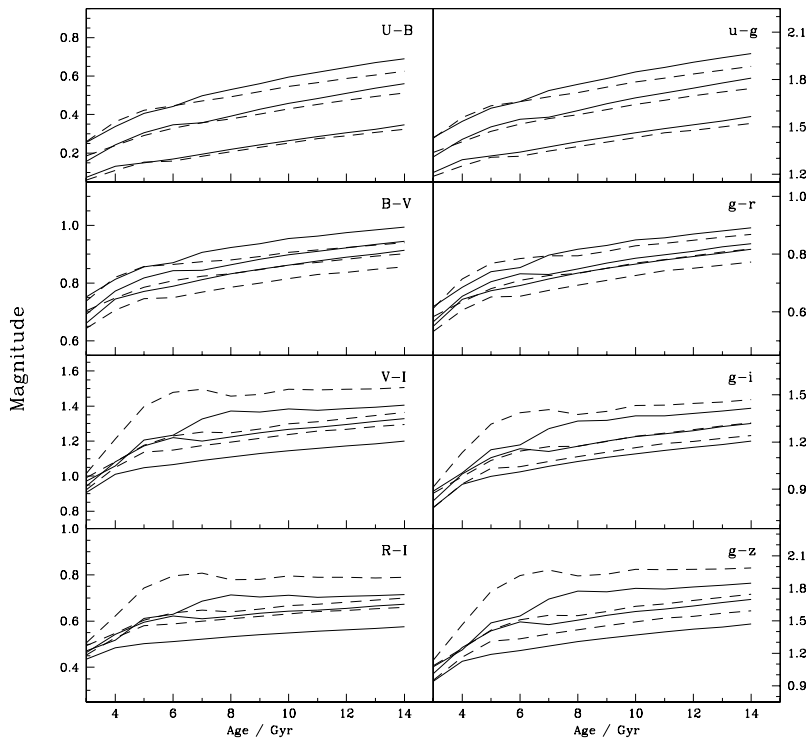}
\caption{The colour evolution predicted by the {\it consistent} models, as described in Section \ref{popsyn}.
The colours are 
indicated in each panel: the left-hand panels show the colours in Johnson-Cousins    
system, and the right-hand panels show the SDSS colours. The solid lines are 
the 
solar-scaled models and the dashed lines are $\alpha$-enhanced models. The 
[Fe/H] 
increases from bottom to top and has values $-$0.5, 0.0 and 0.2. }
\label{f:colours}
\end{center}
\end{figure*}

For the first time we can explore the effect of the $\alpha$ enhancement on the 
broad-band colours of SSPs. 
In Fig. \ref{f:colours} is shown the evolution of the SSP colours as a 
function of age, from the {\it consistent} models of the present work. Solar-scaled models are 
shown as thin black lines and $\alpha$-enhanced models are shown in red thick lines. Iron 
abundances 
from bottom to top are [Fe/H] = $-$0.5, 0.0 and 0.2. 
We can see that for the bluest colours, U$-$B, B$-$V and u$-$g, the $\alpha$-enhanced 
model is slightly bluer than its solar-scaled counterpart, even though the 
total 
metallicity increases. This goes in the opposite direction of what is normally 
expected, i.e. that more metal-rich populations are redder. Given that the 
$\alpha$-enhanced 
isochrones are indeed redder than the solar-scaled ones at the same iron 
abundances (cf. Fig. \ref{fig_coverage}), this effect is coming 
from the stellar energy distributions. The $\alpha$ enhancement changes the 
continuum of a high-metallicity star in a peculiar way, thus the 
$\alpha$-enhanced 
models are bluer in the blue bands, and redder in the red bands than the 
solar-scaled 
models. These findings are in agreement with the work by 
\citet{cassisi+04} 
on stellar colour transformations, and implies that the colour relations 
for metal rich $\alpha$-enhanced populations cannot be described as rescaling 
of the solar-scaled ones. 

\subsection{Evolutionary {\it versus} Spectral effect}\label{evolspec}
With the present models we can verify if the influence of
the $\alpha$ enhancement on spectral observables is dominated by evolution 
(i.e. by the $\alpha$-enhanced tracks) or by stellar spectra 
(i.e. by the $\alpha$-enhanced stars).
In order to quantify each of these effects, we computed two additional 
set of models: the {\it evolution effect} is shown by models in which
$\alpha$-enhanced tracks were combined with 
the solar-scaled stars; similarly, the {\it spectral effect} is shown by models in which  
$\alpha$-enhanced stars were combined with solar-scaled tracks.

\begin{figure}
\begin{center}
\includegraphics[width=8.8cm]{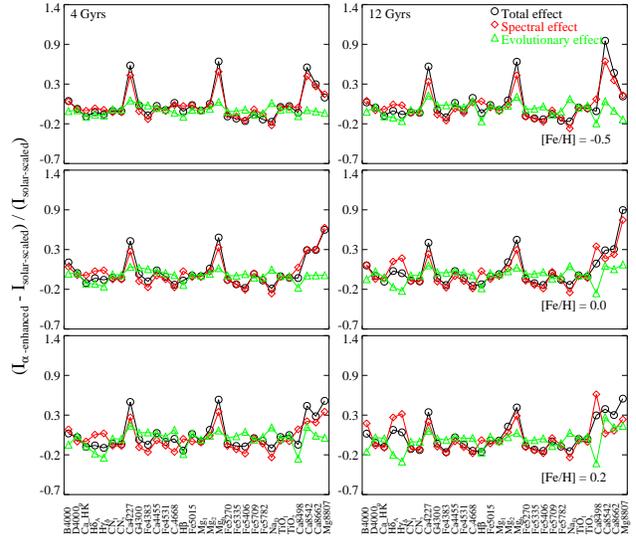}
\caption{
 Differences between the $\alpha$-enhanced predictions
and the solar-scaled ones, in units of the solar-scaled value, for several
spectral indices shown on {\it x}-axis.
Each row corresponds to the prediction of a different [Fe/H] of our models. Left and
righ columns show the predictions at the ages of 4 Gyr and 12 Gyr, respectively.
The green and red lines correspond to the evolutionary effect and spectral effects 
respectively.  The black line are the predictions when both effects are considered 
together. To avoid division by zero, the scales of the 3 Balmer indices and Ca 8498
(which can reach negative values) were shifted arbitrarily upwards.}
\label{f:evolspec_idx}
\end{center}
\end{figure}

\begin{figure}
\begin{center}
\includegraphics[width=8.8cm]{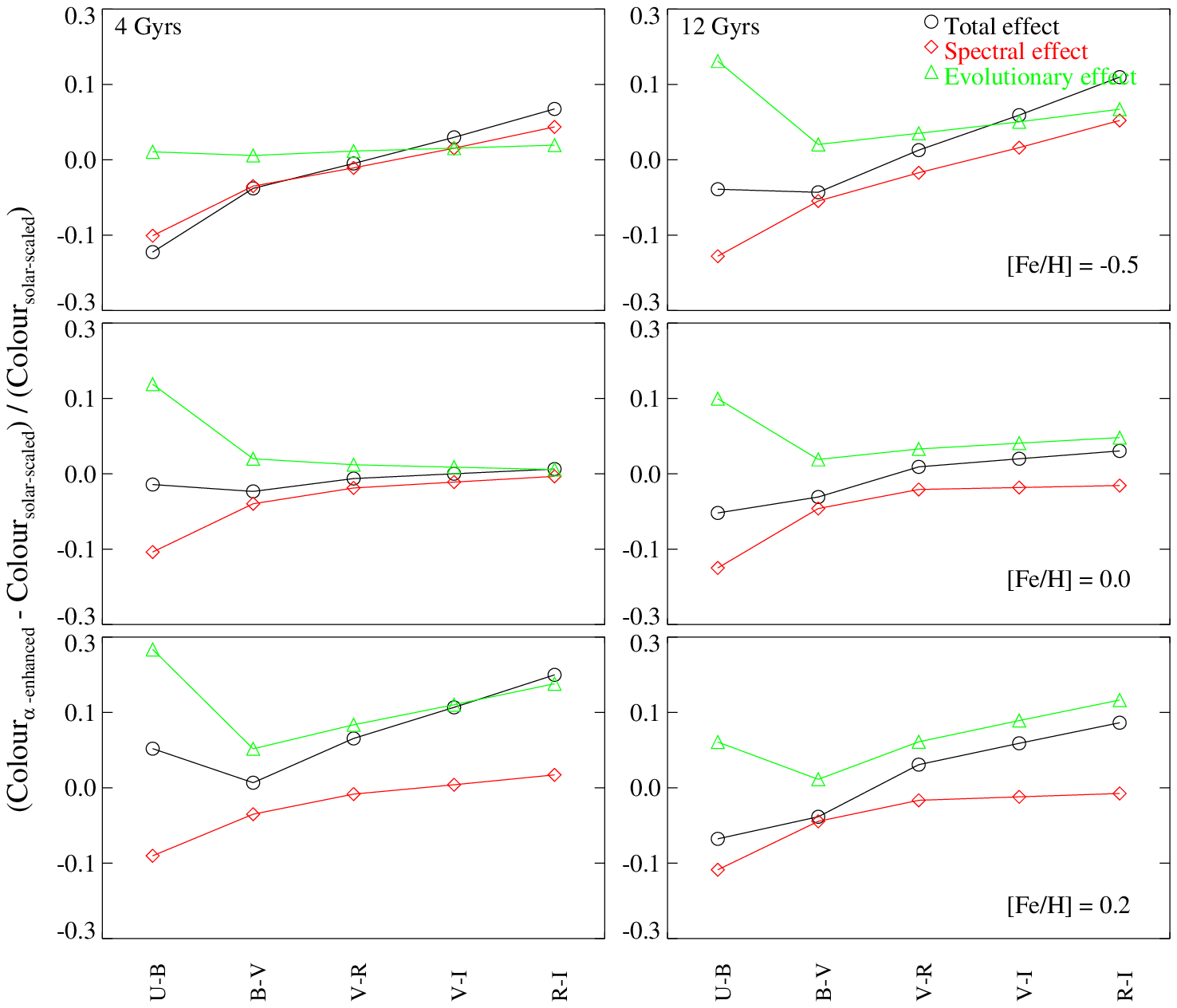}
\caption{Differences between the $\alpha$-enhanced predictions
and the solar-scaled ones, in units of the solar-scaled value, for broad-band colours 
shown on {\it x}-axis. 
Each row corresponds to the prediction of each [Fe/H] of our models. Left and
righ columns show the predictions for 4 Gyr and 12 Gyr respectively.
The green and red lines correspond to the evolutionary effect and spectral effects respectively. 
The
black line are the predictions when both effects are considered together.
}
\label{f:evolspec_colours}
\end{center}
\end{figure}

Our results are presented in Figs \ref{f:evolspec_idx} and \ref{f:evolspec_colours},
for spectral indices and broad-band colours, respectively  (we also provide
tables with model predictions in the Appendix A). In 
each figure, the {\it y}-axis shows the difference between the {\it alpha}-enhanced 
prediction and the solar-scaled one, in units of the solar-scaled value, while 
the {\it x}-axis shows the observables, in order of increasing wavelength.
We present the results for each [Fe/H] of our models (in rows), and for ages
4 and 12 Gyr (in columns).
The green and red lines correspond to the evolutionary effect and spectral effects, 
respectively. The
black lines are the predictions in which both effects are considered, given
by the {\it consistent} models.

In Fig. \ref{f:evolspec_idx} we see that
most of the indices are dominated by spectral effects. Nevertheless there 
are cases where the evolutionary effect is non-negligible, 
namely the Balmer lines.
In general indices at longer wavelengths than Fe5782
are more sensitive to evolution effects.
In the case of colours (Fig. \ref{f:evolspec_colours}), both effects are often
equally important. We note in particular how the spectral effect tends to lower
the U$-$B values, while the evolution
effect tends to increase them. These results highlight the importance of a consistent
modelling of both ingredients.

\subsection{Comparison with SLOAN elliptical galaxies}\label{modidx}

\begin{figure*}
\begin{center}
\includegraphics[width=14cm]{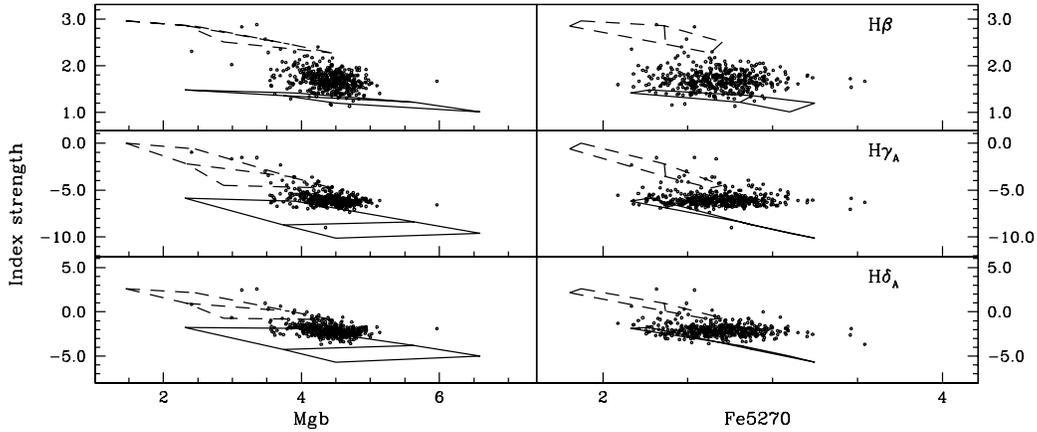}
\caption{Three of the indices defined to measure the Balmer lines, as a 
function 
of Mg $b$ and Fe5270. Solid grid lines are models at fixed age 12 Gyr, 
dashed lines indicate models of 3 Gyr. 
The models with [$\alpha$/Fe] = 0.4 are represented with the right-hand side line, 
i.e., the ones with higher Mg $b$ values. The three parallel lines connect 
fixed [Fe/H] values ($-0.5$, 0.0 and 0.2), and indices become weaker with increasing Fe abundances.
Dots are massive early-type galaxies selected from the 
SDSS-DR4.}
\label{fig:Balmer}
\end{center}
\end{figure*}

\begin{figure*}
\begin{center}
\includegraphics[width=14cm]{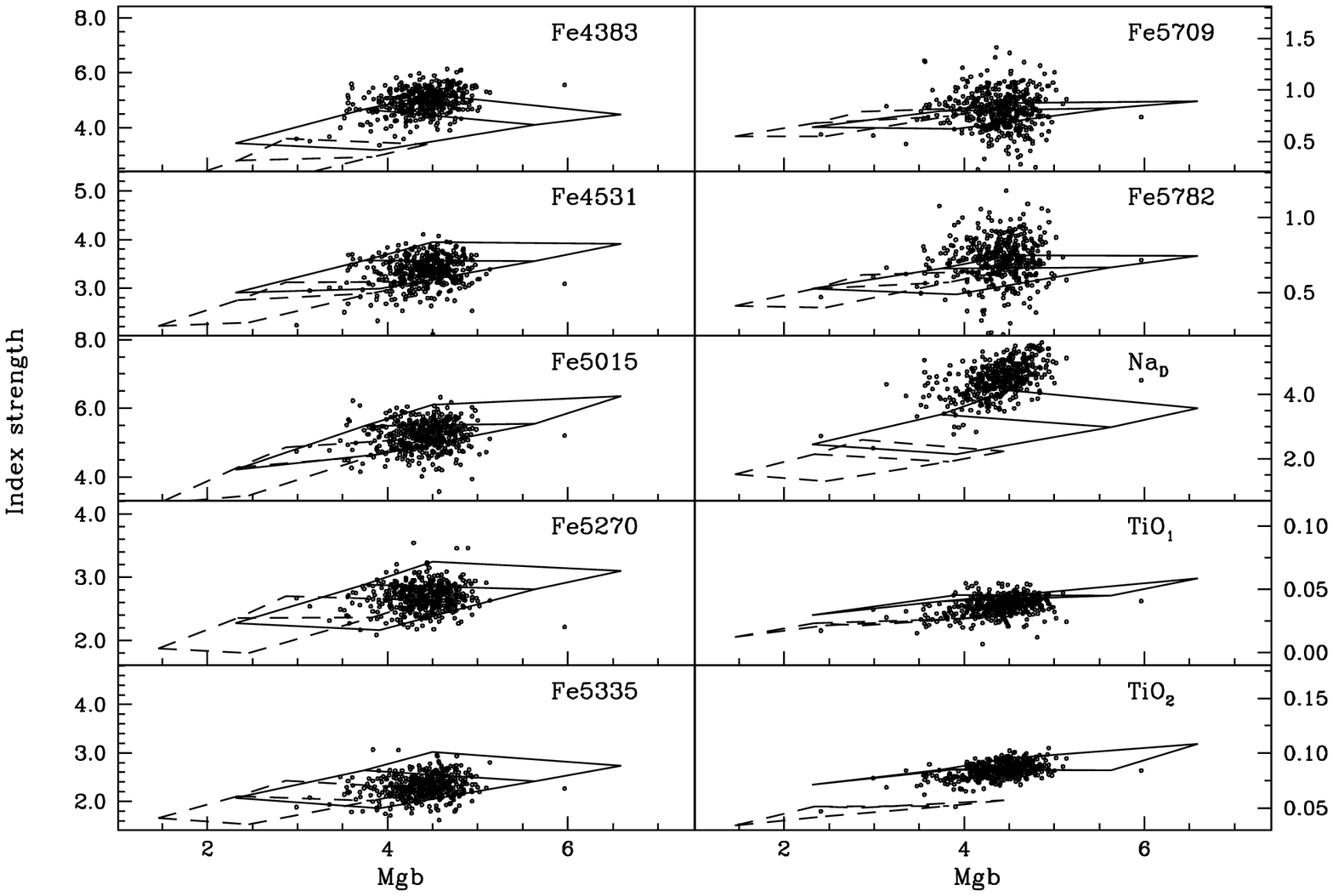}
\caption{Indices highly sensitive to Fe abundances versus Mg $b$. 
TiO$_1$, TiO$_2$ and Na D are also shown here. Models are 
shown as explained in Figure \ref{fig:Balmer}. }
\label{fig:FePeak}
\end{center}
\end{figure*}

\begin{figure*}
\begin{center}
\includegraphics[width=14cm]{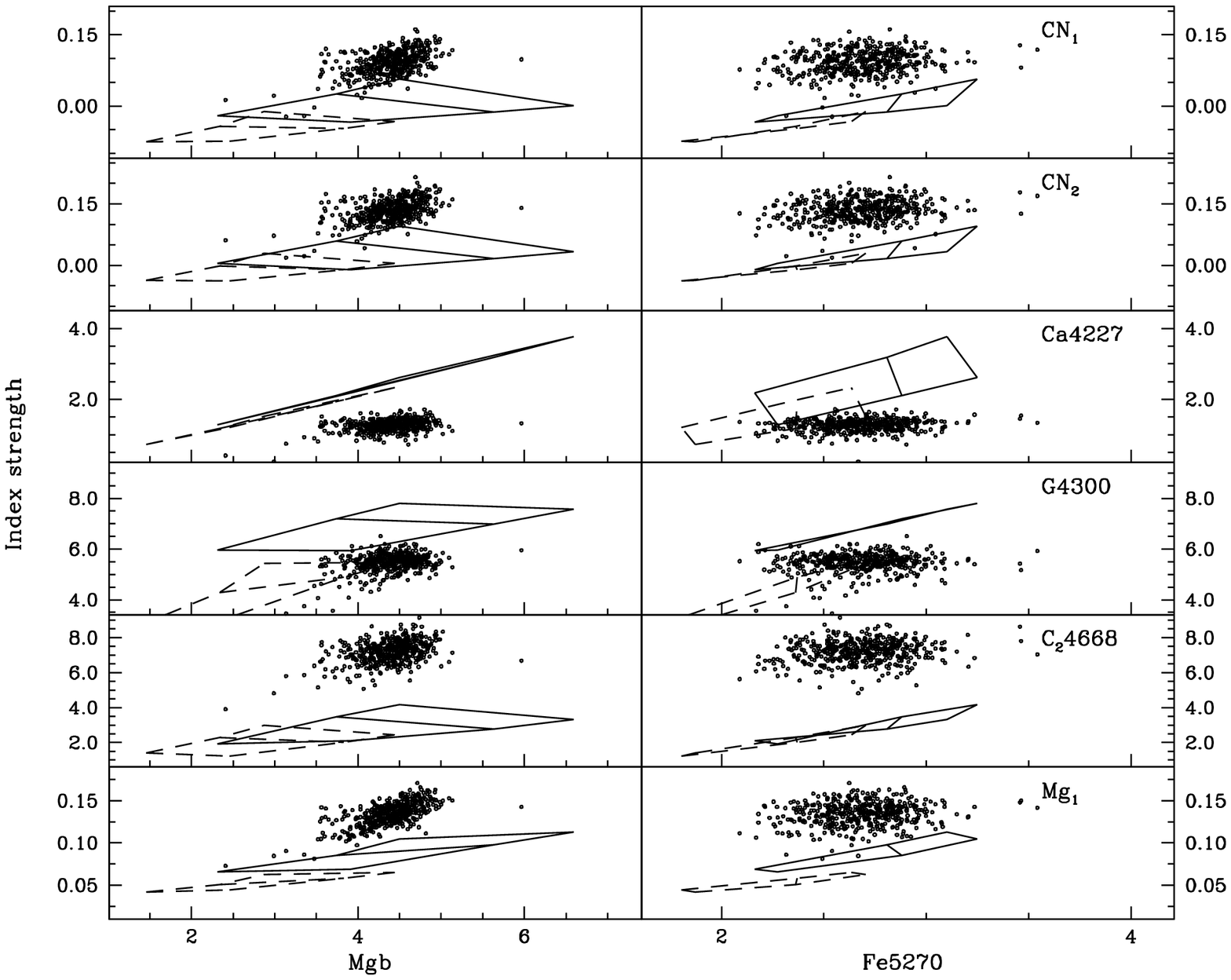}
\caption{Indices that are known to be sensitive to either C or N abundance 
ratios. 
Models are shown as explained in Figure \ref{fig:Balmer}. 
The $\alpha$-enhanced models are the ones with higher Mg $b$ values. 
The [Fe/H] values increase for stronger indices.  }
\label{fig:CNindices}
\end{center}
\end{figure*}

\begin{figure*}
\begin{center}
\includegraphics[width=14cm]{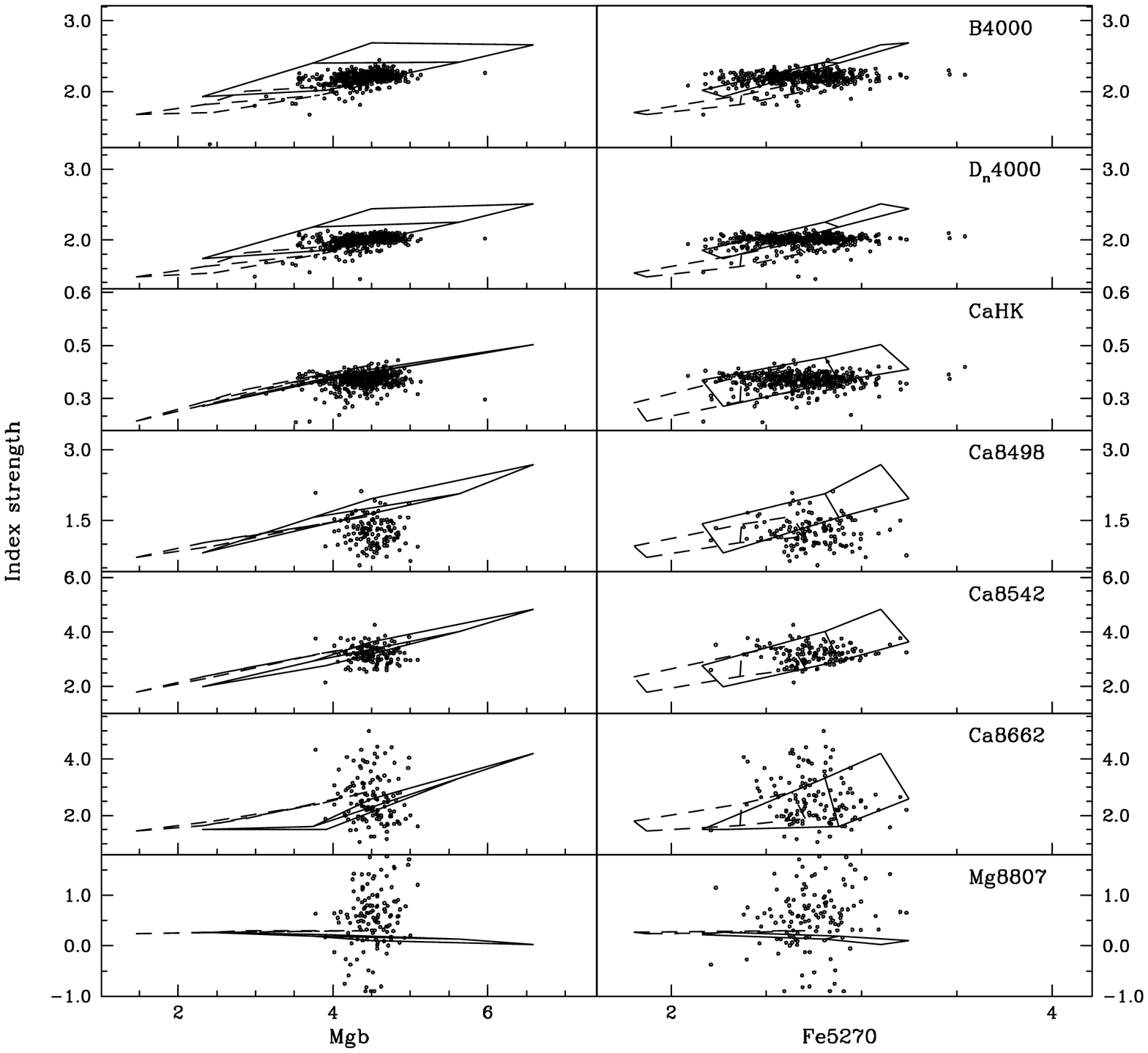}
\caption{Non-Lick indices explored in the present work, as a function of Mg $b$ 
and Fe5270 values. Models are shown as explained in Figure \ref{fig:Balmer}.}
\label{fig:NonLick}
\end{center}
\end{figure*}

In order to further compare our models with observations, we selected 
galaxies from the 
SDSS-DR4 with velocity dispersions 
between 250 and 300 km s$^{-1}$, and concentration parameter c $>$ 2.8. Thus our
sample is dominated by massive early-type galaxies [there might be
some contamination by non-elliptical galaxies, but it should be small.
(see e.g. \citealt{gallazzi+05})].
Only spectra with S/N $\ge$ 40 were considered. 
The models were convolved to the instrumental resolution of the SDSS 
spectrograph, 
and then broadened to a velocity dispersion of 275 km s$^{-1}$.
We compare our models to these observations in {\it index versus index} 
plots, 
shown in Figs \ref{fig:Balmer} to \ref{fig:NonLick}. Models for 
12 Gyr are shown as solid lines grid, and models for 3 Gyr are shown as 
dashed lines. 
The grid lines connect models at fixed [$\alpha$/Fe] (0.0 and 0.4) and at 
fixed [Fe/H] ($-$0.5, 0.0, 0.2).

{\it The Balmer indices:}
Fig. \ref{fig:Balmer} presents the Balmer indices versus an index sensitive 
to Mg (Mg $b$, left-hand column of the figure) and an index sensitive to Fe 
(Fe5270, right-hand column). No attempt was made in calibrating the models, 
due to the uncertainties that affect the synthetic Balmer lines 
(cf. explained in Section \ref{synsedidx}), and we expected some systematic
deviations, but in overall the models 
reproduce very well the observations. Our models show that
H$\delta _A$ and H$\gamma _A$ are more sensitive to $\alpha$ enhancement 
than H$\beta$, confirming the findings by \citet{TMK04} from an independent 
method.

{\it Iron sensitive indices, Na D and TiO:}
In Fig. \ref{fig:FePeak} we present the indices that are highly sensitive 
to iron abundance, besides Na D and TiO, as a function of Mg $b$. 
All the indices sensitive to Fe are very well reproduced, even though 
we note that the bluest index, Fe4383, tend to concentrate around models with lower 
[$\alpha$/Fe] than the other Fe indices. We note from \citet{TB95} work that this 
index is more sensitive to C and Ca abundances, 
which might explain the slightly different behaviour. 
The modelled Na D is weaker than the indices observed in the elliptical 
galaxies. There is strong observational evidence in the literature that Na is enhanced
in ellipticals \citep[e.g.][]{worthey98}. Besides,
the recent work by Lecureur et al. (2007) found a significant overabundance 
of Na/Fe in bulge stars of our Galaxy, which supports the same
findings for ellipticals (given that it 
is reasonable to assume that our bulge and elliptical galaxies share 
a similar star formation history). 
Regarding the TiO indices, particularly difficult to model and extremely 
sensitive to the 
temperature of the giant branch, it is very reassuring that the general 
locus of the models coincides well with the galaxies.

{\it Indices sensitive to C an N abundances:}
Fig. \ref{fig:CNindices} presents all the indices that are known to depend 
either on C or N abundances, or both. Despite of their names, Ca4227 and 
Mg$_1$ were also included in this figure because 
the index Ca4227 has a sensitivity to
C variations of the same order of the sensitivity to Ca abundance variations 
(this is due to lines from the G-band 
CH A$^2$$\Delta$-X$^2$$\Pi$ and blue CN B$^2$$\Sigma$-X$^2$$\Sigma$). 
Concerning Mg$_1$, this index sensitivity to C abundances is even larger than 
its sensitivity to Mg abundances (due to the C$_2$(0,0) band head 
of the Swan A$^3$$\Pi$-X$^3$$\Pi$ system at 5165{\rm \AA}).
In Fig. \ref{fig:CNindices}, the models are poorly representing the elliptical 
galaxy measurements. This is expected, because the present models do not 
take into account variations in the elements that dominate these indices. 
A modelling of C and N variations is required, considering both the 
CN mixing in giants (cf. explained in Section \ref{synsedidx}), and also non-solar 
primordial ratios. 

{\it Non-Lick indices:}
The indices that are not in Lick/IDS system 
are presented in Fig. \ref{fig:NonLick}, versus Mg $b$ (left-hand column) 
and Fe5270 (right-hand column). 
In the first two rows, we show for the first time 
the sensitivity of the 4000 break indices with [$\alpha$/Fe]. In the 
panels showing the 4000 break indices versus Mg $b$, we clearly see that 
the ellipticals concentrate over the $\alpha$-enhanced models. 
Next in the figure we show four Ca indices (the blue 
Ca H\&K and the three near-IR indices). The redder indices show less 
data points because these indices could be measured only in low-redshift galaxies. 
It has been claimed that ellipticals are under abundant in Ca 
\citep[e.g.][]{thomas+03c}, although there is still room for debate 
\citep[e.g.][]{prochaska+05}. In our bulge, which is believed to be 
a reference population for ellipticals, high-resolution stellar spectroscopy 
also suggests low values of [Ca/Fe] \citep[e.g.][]{zoccali+04,alvesbrito+06}, 
while in 
the metal-poor population of our Galaxy the Ca is overabundant, following the trend 
of the other light elements \citep[e.g.][]{cayrel+04}. 
In the present models, the Ca abundance is locked to the other $\alpha$ elements. 
The possibility of using the red Ca indices to measure the [Ca/Fe] of ellipticals 
is worth to be investigated, specially given the fact the blue Lick/IDS 
index sensitive to Ca, Ca4227, responds in non-negligible ways to C 
variations \citep[see also][]{prochaska+05}. Among the four Ca indices, 
Ca8542 is the one that appears to be better reproduced by the models. 
Ca8662 shows a large spread in the observational measurements, spreading around the 
mean values of the models (Ca8662 versus Mg $b$ panel), and spanning the locus 
between the solar-scaled and the $\alpha$-enhanced models (Ca8662 versus Fe5270 panel).
Models for Ca8498 are too strong when compared to observations. 
The models for Ca H\&K, the only blue alternative, are in better agreement 
than the Ca8498, but still marginally too strong.
The galaxy measurements for the last index, Mg8807, show a large spread 
and no unambiguous conclusion can be reached. 

\section{Summary and conclusions}\label{summary}

We have presented a new set of fully synthetic models to compute the 
spectral evolution of stellar populations with solar-scaled or $\alpha$-enhanced
chemical compositions.
These are the first models allowing one to compute the full, high-resolution 
spectral evolution of stellar populations with an abundance pattern different
from that of the solar neighbourhood. 

We focused on consistency while computing and gathering the ingredients for the models. 
As the first ingredient of a synthetic spectral 
model 
are the stellar
tracks, we produced completely new stellar models (cf. Section \ref{stelevol})
in a mass range of
$0.6-10\,\msun$ for six different compositions. Three of them were
for solar-scaled metal ratios and three values of [Fe/H], from subsolar to
supersolar. The remaining three sets had mixtures of identical [Fe/H],
but with $\alpha$-enhanced element ratios, such that total
metallicity, helium and hydrogen abundances are different from the
solar cases. The individual chemical compositions were taken into
account as consistently as possible, most notably in the high- and
low-temperature opacities, for which new up-to-date data were
calculated. The iron abundance values and $\alpha$ elements ratio adopted
are the same as in the synthetic stellar library. 

The stellar synthetic 
library (cf. Section \ref{stelsed}) is based on the one by C05, 
extended to cover very cool giants by computing new high-resolution
spectra based on \citet{plez92} model atmospheres. The photometric
properties of the stellar library was significantly improved by
correcting the effect of missing line opacities, through the comparison
of C05 stars to the flux distributions by
ATLAS9. 
In Section \ref{synsed}, the performance of the synthetic stellar library was analysed by 
comparing 
our models with models built employing empirical 
libraries. The results were much encouraging, 
with broad-band colours and most of the indices being fairly well reproduced. 
The exceptions are spectral indices sensitive to 
either C or N abundances (given that C and N abundance variations
are not yet modelled) and the Balmer lines indices (due to the lack of 
chromosphere models and realistic N-LTE computations).
The remaining differences are consistent with the uncertainties in the 
atmospheric parameters, abundance pattern and flux calibration of the stars 
in the empirical libraries.

With new tracks and an updated C05 library, we
produced {\it consistent} SSP models that include the MS, SGB and RGB
evolutionary phases (cf. Section \ref{popsyn}). These are the most consistent models
available so far that allow for evaluating the influence of 
$\alpha$ enhancement on spectral and photometric observables of stellar 
populations. The differential effect of the $\alpha$ enhancement on
the several spectral and photometric observables mentioned in this work 
are tabulated in the Appendix A.
For the first time the effect of the $\alpha$ enhancement on broad-band colours 
of SSPs can be modelled, and we show that  
the $\alpha$-enhanced colours U$-$B, B$-$V and u$-$g 
are bluer than solar-scaled ones at fixed [Fe/H], in agreement with the 
work by \citet{cassisi+04} on $\alpha$-enhanced colour transformations. 
We intend to look for an observational counterpart to test if our predictions 
are quantitatively correct.

In order to additionally provide a set of models that can be directly compared 
to observations, we extended our evolutionary tracks in a non-consistent way, 
by including 
calculations by \citet{pietrinferni+06} (to account for the HB and
early AGB) and \citet{marigo+07} (to 
account for the TP-AGB).
We show that these spectral models can reproduce the locus 
of ellipticals galaxies in Mg $b$ versus Fe5270 plots (Section \ref{modpred}), where
we see that our models predict slightly stronger Mg indices 
than previous work \citep{TMB03,schiavon07}, for the same value of 
$\alpha$ enhancement.

We also produced models in which we separate the influence of either the evolution 
or the spectra
in the $\alpha$-enhanced predictions.
We conclude that the spectral indices are largely dominated by the spectral influence. Nevertheless
the evolution has non-negligible effects on some indices, among them the Balmer lines.
As for the broad-band colours, the resulting $\alpha$-enhancement effect is a 
combination of both evolution and spectra, and thus any colour predictions 
should include both effects consistently.

We confirm two findings from previous work:
\begin{enumerate}
\item The indices [MgFe], [MgFe]' and [Mg$_2$Fe], defined by \citet{gonzalez93}, 
\citet{TMB03} and BC03 respectively, are fairly insensitive to $\alpha$/Fe. 
We do not confirm the 
same for the index [Mg$_1$Fe] by BC03. No appreciable difference could be 
detected in the behaviour of the three `overall metallicity' indices, for the 
two ages and three values of [Z/H] analysed.
\item The high order Balmer indices H$\delta _A$ and H$\gamma _A$ are more 
sensitive to $\alpha$/Fe variations than H$\beta$, as originally 
shown by \citet{TMK04}.
\end{enumerate}

Lastly, we compared our spectral predictions with indices measured in elliptical 
galaxies observed by the SDSS-DR4. All the indices that are highly 
sensitive to either Fe and $\alpha$ elements are very well matched.  
The Balmer indices are believed to be somewhat underestimated,
nevertheless our models reproduce reasonably well the typical values of massive ellipticals.
Indices that are sensitive to C, N, Ca or Na abundances have to be further modelled
before they can match observed values.

We conclude that our models are suitable for the following applications: 

\begin{enumerate}
\item Deriving the {\it response} of any observable in the wavelength region 
covered by the models to the overall $\alpha$ enhancement. These models are 
an alternative to employing response functions and fitting functions to model 
Lick/IDS indices. More important, they provide a tool to derive the sensitivity 
of several observables, in particular those that are not described by fitting 
functions, e.g. \citet{rose85} indices and broad-band colours.
\item Deriving SSP parameters in a {\it differential} way. As examples, 
the work by \citet{kelson+06} and \citet{smith+06} use 
methods of deriving the parameters differentially, in order to 
avoid the systematic errors present in both models and data.  
\item Full spectrum analysis of stellar populations 
\citep[e.g.][]{cid+05,mathis+06,panter+03}, with the caveat that regions 
known to have important systematic offsets when compared to empirical 
stellar libraries should be avoided in the analysis.
\end{enumerate}

Our models open a new path for stellar population studies, by starting
to fill a gap which existed among the full spectral evolution models. 
We plan to continuously update the models presented here, by including 
future upgrades and extensions of their ingredients.

The spectral models are publicly available upon request.

\vspace{1cm}

{\it Acknowledgments}
 
We are grateful to many that helped this work: P. Weilbacher for the 
discussions that were crucial to develop the flux calibration corrections; 
F. Meissner for performing additional stellar model calculations;  
J. Brinchmann and A. Galazzi for their help with the SDSS data; 
A. Vazdekis, who provided some of his new models prior to publication;
S. Cassisi, for helping us so promptly with the BaSTI tracks; and the
referee, G. Worthey, for his valuable comments. 
PC acknowledges the hospitality at CIDA and IAP institutes, and 
the financial support by: the European Commission's ALFA-II programme 
(LENAC), Fapesp-CNRS process n. 2006/50367-4, 
Fapesp process n. 2005/00397-1, and the European Community under
a Marie Curie International Incoming Fellowship.
BB acknowledges grants from CNPq and Fapesp.

\label{lastpage}
\bibliographystyle{mn2e}

\clearpage
\appendix
\section{Model predictions}

In this Appendix we provide the predictions of the effect of an $\alpha$ enhancement
on broad-band colours and spectral indices. These values were derived from
the {\it consistent} models, as described in Section \ref{popsyn}. 

Table \ref{tabA1} and \ref{tabA2} show the predictions 
for the UBV$_{\rm Johnson}$--RI$_{\rm Cousins}$, and SDSS colours respectively. 
Table \ref{tabA3} to \ref{tabA7} 
show the predictions for the spectral indices, measured on the original 
resolution of the models (FWHM=1 $\angstrom$). 
The first column in the Tables \ref{tabA1} to \ref{tabA7} shows the age in Gyr, and the second 
column shows the [Fe/H] value of the model. From the third column on, each observable 
is related to a pair of columns: the solar-scaled value is shown in column labelled I$_0$, 
and at column $\Delta$ I it is shown the difference between the $\alpha$-enhanced and the solar 
scaled model ($\Delta {\rm I} = {\rm I}_{[\alpha/Fe]=0.4} - {\rm I}_0$).

Tables \ref{tabB1} to \ref{tabB12} show the predictions of evolutionary and spectral effects
separately. The first column lists the observable (colours in the case of Tables \ref{tabB1} to 
\ref{tabB6}, 
indices in the case of Tables \ref{tabB7} to \ref{tabB12}). The second column shows the solar-scaled 
value ${\rm I}_0$. The third and fourth column show the prediction when only the evolutionary or the 
spectral effect is taken into account, respectively. 
The fifth column shows the predictions 
when both effects are considered consistently. 
Six combinations of ages and iron abundances are 
given,
as given in the caption of each table.

\begin{table*}
\caption{Differential effect due to an $\alpha$ enhancement of 0.4 dex on broad-band colours in UBV$_{\rm Johnson}$--RI$_{\rm Cousins}$ system.}
\label{tabA1}
\begin{center}
\begin{tabular}{cccccccccccc}
\hline & & \multicolumn{2}{c}{U$-$B} & \multicolumn{2}{c}{B$-$V} & \multicolumn{2}{c}{V$-$R} & \multicolumn{2}{c}{R$-$I} & \multicolumn{2}{c}{V$-$I} \\
Age (Gyr) & [Fe/H] & I$_0$ & $\Delta$ I & I$_0$ & $\Delta$ I & I$_0$ & $\Delta$ I & I$_0$ & $\Delta$ I & I$_0$ & $\Delta$ I \\
\hline
 3. & $-$0.5 &     0.0744 &    $-$0.0145 &     0.6607 &    $-$0.0173 &     0.4712 &    $-$0.0024 &     0.4346 &     0.0185 &     0.9059 &     0.0160\\
 4. & $-$0.5 &     0.1324 &    $-$0.0219 &     0.7443 &    $-$0.0381 &     0.5264 &    $-$0.0035 &     0.4840 &     0.0441 &     1.0103 &     0.0406\\
 5. & $-$0.5 &     0.1503 &     0.0023 &     0.7709 &    $-$0.0245 &     0.5453 &     0.0127 &     0.5026 &     0.0769 &     1.0479 &     0.0895\\
 6. & $-$0.5 &     0.1687 &    $-$0.0103 &     0.7892 &    $-$0.0398 &     0.5548 &     0.0065 &     0.5107 &     0.0774 &     1.0656 &     0.0838\\
 7. & $-$0.5 &     0.1943 &    $-$0.0103 &     0.8117 &    $-$0.0428 &     0.5672 &     0.0068 &     0.5216 &     0.0784 &     1.0889 &     0.0851\\
 8. & $-$0.5 &     0.2200 &    $-$0.0130 &     0.8317 &    $-$0.0463 &     0.5784 &     0.0064 &     0.5318 &     0.0785 &     1.1101 &     0.0850\\
\hline
\end{tabular}
\end{center}
\end{table*}
 
\begin{table*}
\caption{Differential effect due to an $\alpha$ enhancement of 0.4dex on broad-band colours in the SDSS photometric bands.}
\label{tabA2}
\begin{center}
\begin{tabular}{cccccccccc}
\hline & & \multicolumn{2}{c}{u$-$g} & \multicolumn{2}{c}{g$-$r} & \multicolumn{2}{c}{g$-$i} & \multicolumn{2}{c}{g$-$z} \\
Age (Gyr) & [Fe/H]    & I$_0$ & $\Delta$ I & I$_0$ & $\Delta$ I & I$_0$ & $\Delta$ I & I$_0$ & $\Delta$ I \\
\hline
 3. & $-$0.5 &     1.2111 &    $-$0.0262 &     0.5499 &    $-$0.0176 &     0.7852 &    $-$0.0021 &     0.9399 &     0.0091\\
 4. & $-$0.5 &     1.2917 &    $-$0.0390 &     0.6438 &    $-$0.0377 &     0.9325 &     0.0009 &     1.1262 &     0.0388\\
 5. & $-$0.5 &     1.3157 &    $-$0.0085 &     0.6740 &    $-$0.0224 &     0.9829 &     0.0492 &     1.1912 &     0.1204\\
 6. & $-$0.5 &     1.3396 &    $-$0.0261 &     0.6920 &    $-$0.0380 &     1.0099 &     0.0331 &     1.2241 &     0.1095\\
 7. & $-$0.5 &     1.3729 &    $-$0.0267 &     0.7148 &    $-$0.0397 &     1.0447 &     0.0326 &     1.2673 &     0.1105\\
 8. & $-$0.5 &     1.4060 &    $-$0.0306 &     0.7350 &    $-$0.0421 &     1.0760 &     0.0304 &     1.3064 &     0.1091\\
\hline
\end{tabular}
\end{center}
\end{table*}
 
\begin{landscape}
\begin{table}
\caption{Differential effect due to an $\alpha$ enhancement of 0.4dex on indices  (measured on the resolution of FWHM = 1 $\angstrom$). Indices: B4000 through CN$_2$.}
\label{tabA3}
\begin{center}
\begin{tabular}{cccccccccccccccc}
\hline & & \multicolumn{2}{c}{B4000} & \multicolumn{2}{c}{D$_n$(4000)}   & \multicolumn{2}{c}{Ca H\&K} & \multicolumn{2}{c}{H$\delta _A$} & \multicolumn{2}{c}{H$\gamma _A$}  & \multicolumn{2}{c}{CN$_1$}  & \multicolumn{2}{c}{CN$_2$} \\
Age (Gyr) &  [Fe/H]   &I$_0$ & $\Delta$ I &I$_0$ & $\Delta$ I &I$_0$ & $\Delta$ I &I$_0$ & $\Delta$ I &I$_0$ & $\Delta$ I &I$_0$ & $\Delta$ I &I$_0$ & $\Delta$ I \\
\hline
 3. & $-$0.5 &     1.6008 &     0.0323 &     1.4143 &     0.0626 &     0.2291 &     0.0539 &     3.4298 &    $-$0.6214 &     1.5947 &    $-$0.9528 &    $-$0.0762 &     0.0069 &    $-$0.0255 &     0.0030\\
 4. & $-$0.5 &     1.6586 &     0.0377 &     1.4793 &     0.0661 &     0.2410 &     0.0586 &     1.6118 &    $-$0.2540 &    $-$0.6550 &    $-$0.5858 &    $-$0.0466 &    $-$0.0020 &    $-$0.0027 &    $-$0.0047\\
 5. & $-$0.5 &     1.6804 &     0.0748 &     1.5059 &     0.1006 &     0.2459 &     0.0671 &     0.8341 &    $-$0.4765 &    $-$1.6188 &    $-$0.9481 &    $-$0.0359 &    $-$0.0004 &     0.0048 &    $-$0.0040\\
 6. & $-$0.5 &     1.7039 &     0.0658 &     1.5321 &     0.0921 &     0.2508 &     0.0658 &     0.2837 &    $-$0.2765 &    $-$2.3335 &    $-$0.6851 &    $-$0.0290 &    $-$0.0037 &     0.0094 &    $-$0.0068\\
 7. & $-$0.5 &     1.7354 &     0.0695 &     1.5644 &     0.0955 &     0.2565 &     0.0668 &    $-$0.2487 &    $-$0.2008 &    $-$3.0373 &    $-$0.6029 &    $-$0.0222 &    $-$0.0056 &     0.0139 &    $-$0.0085\\
 8. & $-$0.5 &     1.7678 &     0.0694 &     1.5963 &     0.0961 &     0.2618 &     0.0674 &    $-$0.6768 &    $-$0.1587 &    $-$3.6161 &    $-$0.5486 &    $-$0.0172 &    $-$0.0066 &     0.0172 &    $-$0.0095\\
\hline
\end{tabular}
\end{center}
\end{table}
 
\end{landscape}
\begin{landscape}
\begin{table}
\caption{The same as in Table \ref{tabA3}, for indices Ca4227 through H$\beta$.}
\label{tabA4}
\begin{center}
\begin{tabular}{cccccccccccccccc}
\hline & &\multicolumn{2}{c}{Ca4227} & \multicolumn{2}{c}{G4300} & \multicolumn{2}{c}{Fe4383} & \multicolumn{2}{c}{Ca4455} & \multicolumn{2}{c}{Fe4531} & \multicolumn{2}{c}{C$_2$4668}  & \multicolumn{2}{c}{H$\beta$} \\
Age (Gyr) &  [Fe/H]   &I$_0$ & $\Delta$ I &I$_0$ & $\Delta$ I &I$_0$ & $\Delta$ I &I$_0$ & $\Delta$ I &I$_0$ & $\Delta$ I &I$_0$ & $\Delta$ I &I$_0$ & $\Delta$ I \\
\hline
 3. & $-$0.5 &     0.8500 &     0.5602 &     2.7668 &     0.4818 &     1.6918 &    $-$0.0150 &     0.8666 &     0.0796 &     2.3692 &     0.0706 &     1.2345 &    $-$0.0036 &     3.5638 &    $-$0.2165\\
 4. & $-$0.5 &     1.0898 &     0.6722 &     3.8499 &     0.3030 &     2.4548 &    $-$0.1938 &     1.0320 &     0.0777 &     2.6703 &    $-$0.0015 &     1.5070 &     0.1675 &     2.7801 &    $-$0.0430\\
 5. & $-$0.5 &     1.2181 &     0.8224 &     4.3163 &     0.4910 &     2.7271 &    $-$0.0867 &     1.0914 &     0.1392 &     2.7715 &     0.0560 &     1.5713 &     0.3954 &     2.4842 &    $-$0.1414\\
 6. & $-$0.5 &     1.3017 &     0.8238 &     4.6837 &     0.3541 &     2.9253 &    $-$0.2107 &     1.1375 &     0.1213 &     2.8500 &     0.0167 &     1.6233 &     0.3478 &     2.2824 &    $-$0.0677\\
 7. & $-$0.5 &     1.3945 &     0.8738 &     5.0497 &     0.3014 &     3.1403 &    $-$0.2511 &     1.1923 &     0.1287 &     2.9417 &     0.0069 &     1.7081 &     0.3578 &     2.0881 &    $-$0.0584\\
 8. & $-$0.5 &     1.4809 &     0.9159 &     5.3571 &     0.2566 &     3.3236 &    $-$0.2888 &     1.2424 &     0.1329 &     3.0233 &    $-$0.0043 &     1.7936 &     0.3513 &     1.9349 &    $-$0.0584\\
\hline
\end{tabular}
\end{center}
\end{table}
 
\end{landscape}
\begin{landscape}
\begin{table}
\caption{The same as in Table \ref{tabA3}, for indices Fe5015 through Fe5406.}
\label{tabA5}
\begin{center}
\begin{tabular}{cccccccccccccccc}
\hline & &\multicolumn{2}{c}{Fe5015} &  \multicolumn{2}{c}{Mg$_1$}  & \multicolumn{2}{c}{Mg$_2$}  & \multicolumn{2}{c}{Mg$_b$} & \multicolumn{2}{c}{Fe5270} & \multicolumn{2}{c}{Fe5335} & \multicolumn{2}{c}{Fe5406} \\
Age (Gyr) &  [Fe/H]   &I$_0$ & $\Delta$ I &I$_0$ & $\Delta$ I &I$_0$ & $\Delta$ I &I$_0$ & $\Delta$ I &I$_0$ & $\Delta$ I &I$_0$ & $\Delta$ I &I$_0$ & $\Delta$ I \\
\hline
 3. & $-$0.5 &     3.8412 &     0.2188 &     0.0421 &     0.0022 &     0.1204 &     0.0321 &     1.5143 &     1.0552 &     2.2184 &    $-$0.1135 &     2.3161 &    $-$0.2041 &     1.2569 &    $-$0.1683\\
 4. & $-$0.5 &     4.3847 &     0.2929 &     0.0527 &     0.0002 &     0.1424 &     0.0364 &     1.7807 &     1.2532 &     2.4447 &    $-$0.1911 &     2.5867 &    $-$0.2710 &     1.4855 &    $-$0.2037\\
 5. & $-$0.5 &     4.5762 &     0.4993 &     0.0553 &     0.0021 &     0.1514 &     0.0448 &     1.9406 &     1.4656 &     2.5020 &    $-$0.1595 &     2.6611 &    $-$0.2270 &     1.5601 &    $-$0.1598\\
 6. & $-$0.5 &     4.6960 &     0.4162 &     0.0568 &    $-$0.0006 &     0.1571 &     0.0421 &     2.0610 &     1.5064 &     2.5404 &    $-$0.2025 &     2.7032 &    $-$0.2842 &     1.6014 &    $-$0.2011\\
 7. & $-$0.5 &     4.8493 &     0.4112 &     0.0594 &     0.0001 &     0.1642 &     0.0445 &     2.1821 &     1.5752 &     2.6000 &    $-$0.2047 &     2.7682 &    $-$0.2919 &     1.6595 &    $-$0.2053\\
 8. & $-$0.5 &     4.9913 &     0.3941 &     0.0620 &     0.0004 &     0.1710 &     0.0463 &     2.2935 &     1.6324 &     2.6565 &    $-$0.2112 &     2.8288 &    $-$0.3033 &     1.7136 &    $-$0.2127\\
\hline
\end{tabular}
\end{center}
\end{table}
 
\end{landscape}
\begin{landscape}
\begin{table}
\caption{The same as in Table \ref{tabA3}, for indices Fe5709 through TiO$_2$.}
\label{tabA6}
\begin{center}
\begin{tabular}{cccccccccccccccc}
\hline & &\multicolumn{2}{c}{Fe5709} & \multicolumn{2}{c}{Fe5782} & \multicolumn{2}{c}{Na$_D$} & \multicolumn{2}{c}{TiO$_1$}  & \multicolumn{2}{c}{TiO$_2$} \\
Age (Gyr) &  [Fe/H]   &I$_0$ & $\Delta$ I &I$_0$ & $\Delta$ I &I$_0$ & $\Delta$ I &I$_0$ & $\Delta$ I &I$_0$ & $\Delta$ I \\
\hline
 3. & $-$0.5 &     0.5996 &    $-$0.0107 &     0.5385 &    $-$0.0524 &     1.7637 &    $-$0.2208 &     0.0128 &     0.0108 &     0.0368 &     0.0118\\
 4. & $-$0.5 &     0.6780 &    $-$0.0380 &     0.6265 &    $-$0.0774 &     2.1418 &    $-$0.3164 &     0.0185 &     0.0202 &     0.0512 &     0.0246\\
 5. & $-$0.5 &     0.6877 &    $-$0.0209 &     0.6430 &    $-$0.0654 &     2.3441 &    $-$0.3303 &     0.0218 &     0.0242 &     0.0590 &     0.0299\\
 6. & $-$0.5 &     0.6932 &    $-$0.0399 &     0.6496 &    $-$0.0903 &     2.4526 &    $-$0.3737 &     0.0233 &     0.0217 &     0.0621 &     0.0249\\
 7. & $-$0.5 &     0.7063 &    $-$0.0388 &     0.6655 &    $-$0.0932 &     2.5739 &    $-$0.3918 &     0.0253 &     0.0209 &     0.0664 &     0.0228\\
 8. & $-$0.5 &     0.7179 &    $-$0.0394 &     0.6803 &    $-$0.0976 &     2.6871 &    $-$0.4113 &     0.0275 &     0.0197 &     0.0708 &     0.0201\\
\hline
\end{tabular}
\end{center}
\end{table}
 
\end{landscape}
\begin{landscape}
\begin{table}
\caption{The same as in Table \ref{tabA3}, for indices Ca8498 through Mg8807.}
\label{tabA7}
\begin{center}
\begin{tabular}{cccccccccccccccc}
\hline & &\multicolumn{2}{c}{Ca8498}   &\multicolumn{2}{c}{Ca8542}	 & \multicolumn{2}{c}{Ca8662}	 & \multicolumn{2}{c}{Mg8807}	  \\
Age (Gyr) &  [Fe/H]   &I$_0$ & $\Delta$ I &I$_0$ & $\Delta$ I &I$_0$ & $\Delta$ I &I$_0$ & $\Delta$ I \\
\hline
 3. & $-$0.5 &     0.7366 &     0.3072 &     2.0579 &     0.6445 &     1.7425 &     0.4107 &     0.4466 &     0.0639\\
 4. & $-$0.5 &     0.7456 &     0.4467 &     2.1036 &     0.7198 &     1.7658 &     0.2994 &     0.4704 &     0.0561\\
 5. & $-$0.5 &     0.7509 &     0.6195 &     2.1108 &     0.8734 &     1.7559 &     0.2480 &     0.4772 &     0.0541\\
 6. & $-$0.5 &     0.7643 &     0.7000 &     2.1356 &     0.9641 &     1.7636 &     0.2643 &     0.4833 &     0.0506\\
 7. & $-$0.5 &     0.7756 &     0.7238 &     2.1541 &     0.9935 &     1.7662 &     0.2763 &     0.4900 &     0.0510\\
 8. & $-$0.5 &     0.7859 &     0.7427 &     2.1678 &     1.0195 &     1.7649 &     0.2895 &     0.4957 &     0.0514\\
\hline
\end{tabular}
\end{center}
\end{table}
\end{landscape}
 
\begin{table}
\caption{Separated effect of the $\alpha$ enhancement on broad-band colours due to evolutionary
or spectral effects, for an SSP with 4 Gyr and [Fe/H] = $-$0.5. }
\label{tabB1}
\begin{center}
\begin{tabular}{lcccc}
\hline 
Colour & I$_{0}$ & ${\rm I}_{evolution}$ & ${\rm I}_{spectra}$ & I\\
\hline
U$-$B  &     0.1324 &    0.1343 &    0.1144 &    0.1105 \\
B$-$V  &     0.7443 &    0.7501 &    0.7092 &    0.7062 \\
V$-$R  &     0.5264 &    0.5346 &    0.5188 &    0.5229 \\
V$-$I  &     1.0103 &    1.0313 &    1.0313 &    1.0509 \\
R$-$I  &     0.4840 &    0.4967 &    0.5125 &    0.5281 \\
\end{tabular} 
\end{center}
\end{table}

\begin{table}
\caption{Separated effect of the $\alpha$ enhancement on broad-band colours due to evolutionary
or spectral effects, for an SSP with 12 Gyr and [Fe/H] = $-$0.5. }
\label{tabB2}
\begin{center}
\begin{tabular}{lcccc}
\hline 
Colour & I$_{0}$ & ${\rm I}_{evolution}$ & ${\rm I}_{spectra}$ & I\\
\hline
U$-$B  &        0.3052  &  0.3592 &  0.2525  &  0.2891  \\
B$-$V  &        0.8891  &  0.9136 &  0.8234  &  0.8374  \\
V$-$R  &        0.6107  &  0.6397 &  0.5965  &  0.6213  \\
V$-$I  &        1.1730  &  1.2529 &  1.1985  &  1.2669  \\
R$-$I  &        0.5624  &  0.6133 &  0.6020  &  0.6457  \\
\end{tabular} 
\end{center}
\end{table}

\begin{table}
\caption{Separated effect of the $\alpha$ enhancement on broad-band colours due to evolutionary
or spectral effects, for an SSP with 4 Gyr and [Fe/H] = 0.0. }
\label{tabB3}
\begin{center}
\begin{tabular}{lcccc}
\hline 
Colour & I$_{0}$ & ${\rm I}_{evolution}$ & ${\rm I}_{spectra}$ & I\\
\hline
U$-$B  &        0.2431 &   0.2821 &   0.2090  &  0.2384  \\
B$-$V  &        0.7726 &   0.7933 &   0.7311  &  0.7482  \\
V$-$R  &        0.5396 &   0.5483 &   0.5259  &  0.5350  \\
V$-$I  &        1.0826 &   1.0954 &   1.0664  &  1.0826  \\
R$-$I  &        0.5430 &   0.5471 &   0.5405  &  0.5476  \\
\end{tabular} 
\end{center}
\end{table}

\begin{table}
\caption{Separated effect of the $\alpha$ enhancement on broad-band colours due to evolutionary
or spectral effects, for an SSP with 12 Gyr and [Fe/H] = 0.0. }
\label{tabB4}
\begin{center}
\begin{tabular}{lcccc}
\hline 
Colour & I$_{0}$ & ${\rm I}_{evolution}$ & ${\rm I}_{spectra}$ & I\\
\hline
U$-$B  &        0.5098   & 0.5784  &  0.4238   &  0.4740  \\
B$-$V  &        0.9215   & 0.9452  &  0.8641   &  0.8831  \\
V$-$R  &        0.6388   & 0.6672  &  0.6208   &  0.6467  \\
V$-$I  &        1.2940   & 1.3649  &  1.2622   &  1.3288  \\
R$-$I  &        0.6552   & 0.6977  &  0.6414   &  0.6820  \\
\end{tabular} 
\end{center}
\end{table}

\begin{table}
\caption{Separated effect of the $\alpha$ enhancement on broad-band colours due to evolutionary
or spectral effects, for an SSP with 4 Gyr and [Fe/H] = +0.2. }
\label{tabB5}
\begin{center}
\begin{tabular}{lcccc}
\hline 
Colour & I$_{0}$ & ${\rm I}_{evolution}$ & ${\rm I}_{spectra}$ & I\\
\hline
U$-$B  &        0.3386  &  0.4225  &   0.2973  &   0.3623  \\
B$-$V  &        0.8127  &  0.8694  &   0.7742  &   0.8200  \\
V$-$R  &        0.5415  &  0.6026  &   0.5353  &   0.5893  \\
V$-$I  &        1.0588  &  1.2163  &   1.0645  &   1.2112  \\
R$-$I  &        0.5173  &  0.6137  &   0.5292  &   0.6219  \\
\end{tabular} 
\end{center}
\end{table}

\begin{table}
\caption{Separated effect of the $\alpha$ enhancement on broad-band colours due to evolutionary
or spectral effects, for an SSP with 12 Gyr and [Fe/H] = +0.2. }
\label{tabB6}
\begin{center}
\begin{tabular}{lcccc}
\hline 
Colour & I$_{0}$ & ${\rm I}_{evolution}$ & ${\rm I}_{spectra}$ & I\\
\hline
U$-$B  &       0.6456 &    0.6984  &  0.5507  &   0.5865  \\
B$-$V  &       0.9743 &    0.9890  &  0.9156  &   0.9236  \\
V$-$R  &       0.6786 &    0.7344  &  0.6633  &   0.7067  \\
V$-$I  &       1.3857 &    1.5525  &  1.3630  &   1.4960  \\
R$-$I  &       0.7070 &    0.8181  &  0.6997  &   0.7893  \\
\end{tabular} 
\end{center}
\end{table}

\begin{table}
\caption{Separated effect of the $\alpha$ enhancement on spectral indices due to evolutionary
or spectral effects, for an SSP with 4 Gyr and [Fe/H] = $-$0.5. }
\label{tabB7}
\begin{center}
\begin{tabular}{lcccc}
\hline 
Spectral index & I$_0$ & ${\rm I}_{evolution}$ & ${\rm I}_{spectra}$ & I\\
\hline
 B4000          &   0.4539 &   0.4505 &   0.5085 &    0.5113\\
 D4000$_{vn}$   &   1.6582 &   1.6646 &   1.6908 &    1.6959\\
 Ca$_{HK}$      &   2.7196 &   2.5003 &   2.7126 &    2.5062\\
 H$_{\delta A}$ &  11.5626 &  10.9137 &  11.9172 &    11.311\\
 H$_{\gamma A}$ &  11.2905 &  10.5411 &  11.3991 &    10.718\\
 CN$_1$         &   0.8885 &   0.9034 &   0.8685 &    0.8843\\
 CN$_2$         &   0.9883 &   0.9990 &   0.9658 &    0.9779\\
\end{tabular}
\end{center}
\end{table}

\begin{table}
\caption{Separated effect of the $\alpha$ enhancement on spectral indices due to evolutionary
or spectral effects, for an SSP with 12 Gyr and [Fe/H] = $-$0.5. }
\label{tabB8}
\begin{center}
\begin{tabular}{lcccc}
\hline 
Spectral index & I$_0$ & ${\rm I}_{evolution}$ & ${\rm I}_{spectra}$ & I\\
\hline
 B4000          &     0.4936   &    0.4859   &   0.5577   &   0.5466  \\
 D4000$_{vn}$   &     1.8816   &    1.9691   &   1.8880   &   1.9554  \\
 Ca$_{HK}$      &     1.7882   &    1.6494   &   1.8159   &   1.6669  \\
 H$_{\delta A}$ &     8.0322   &    7.2890   &   8.6714   &   8.0117  \\
 H$_{\gamma A}$ &     6.6842   &    5.7513   &   7.1545   &   6.3262  \\
 CN$_1$         &     0.9758   &    0.9833   &   0.9464   &   0.9527  \\
 CN$_2$         &     1.0534   &    1.0556   &   1.0221   &   1.0225  \\
\end{tabular}
\end{center}
\end{table}

\begin{table}
\caption{Separated effect of the $\alpha$ enhancement on spectral indices due to evolutionary
or spectral effects, for an SSP with 4 Gyr and [Fe/H] = 0.0. }
\label{tabB9}
\begin{center}
\begin{tabular}{lcccc}
\hline 
Spectral index & I$_0$ & ${\rm I}_{evolution}$ & ${\rm I}_{spectra}$ & I\\
\hline
 B4000          &        0.5440     &  0.5617   &    0.6097  &   0.6422 \\
 D4000$_{vn}$   &        1.8493     &  1.9208   &    1.8724  &   1.9236\\
 Ca$_{HK}$      &        2.5881     &  2.3232   &    2.6160  &   2.3533\\
 H$_{\delta A}$ &        9.7499     &  8.7550   &   10.3923  &   9.4794\\
 H$_{\gamma A}$ &        8.7753     &  7.5595   &    9.4472  &   8.3371\\
 CN$_1$         &        0.9517     &  0.9722   &    0.9105  &   0.9321 \\
 CN$_2$         &        1.0621     &  1.0769   &    1.0150  &   1.0308\\
\end{tabular}
\end{center}
\end{table}

\begin{table}
\caption{Separated effect of the $\alpha$ enhancement on spectral indices due to evolutionary
or spectral effects, for an SSP with 12 Gyr and [Fe/H] = 0.0. }
\label{tabB10}
\begin{center}
\begin{tabular}{lcccc}
\hline 
Spectral index & I$_0$ & ${\rm I}_{evolution}$ & ${\rm I}_{spectra}$ & I\\
\hline
 B4000          &        0.5696  &   0.5457 &     0.6531 &    0.6493 \\
 D4000$_{vn}$   &        2.3013  &   2.4416 &     2.2137 &    2.3030\\
 Ca$_{HK}$      &        1.9652  &   1.9285 &     1.9126 &    1.8198\\
 H$_{\delta A}$ &        5.6702  &   4.8528 &     6.7580 &    6.0480\\
 H$_{\gamma A}$ &        4.1399  &   3.3264 &     5.1257 &    4.2950\\
 CN$_1$         &        1.0715  &   1.0829 &     0.9981 &    1.0069\\
 CN$_2$         &        1.1735  &   1.1825 &     1.0871 &    1.0911\\
\end{tabular}
\end{center}
\end{table}

\begin{table}
\caption{Separated effect of the $\alpha$ enhancement on spectral indices due to evolutionary
or spectral effects, for an SSP with 4 Gyr and [Fe/H] = +0.2. }
\label{tabB11}
\begin{center}
\begin{tabular}{lcccc}
\hline 
Spectral index & I$_0$ & ${\rm I}_{evolution}$ & ${\rm I}_{spectra}$ & I\\
\hline
 B4000          &        0.6261   &  0.6004   &  0.7271 &    0.6930   \\
 D4000$_{vn}$   &        2.0146   &  2.1649   &  2.0153 &    2.1314  \\
 Ca$_{HK}$      &        2.4671   &  2.3131   &  2.4733 &    2.2940  \\
 H$_{\delta A}$ &        8.5177   &  7.1759   &  9.3009 &    8.0866  \\
 H$_{\gamma A}$ &        7.2820   &  5.7248   &  8.0730 &    6.6751  \\
 CN$_1$         &        1.0036   &  1.0418   &  0.9501 &    0.9767   \\
 CN$_2$         &        1.1212   &  1.1601   &  1.0585 &    1.0814  \\
\end{tabular}
\end{center}
\end{table}

\begin{table}
\caption{Separated effect of the $\alpha$ enhancement on spectral indices due to evolutionary
or spectral effects, for an SSP with 12 Gyr and [Fe/H] = 0.2. }
\label{tabB12}
\begin{center}
\begin{tabular}{lcccc}
\hline 
Spectral index & I$_0$ & ${\rm I}_{evolution}$ & ${\rm I}_{spectra}$ & I\\
\hline
 B4000          &        0.5726   &  0.4951    &  0.7103  &    0.6324   \\
 D4000$_{vn}$   &        2.5717   &  2.7140    &  2.4423  &    2.5315  \\
 Ca$_{HK}$      &        2.1678   &  2.2592    &  1.9936  &    2.0209  \\
 H$_{\delta A}$ &        4.1421   &  3.4098    &  5.4680  &    4.7830  \\
 H$_{\gamma A}$ &        2.7385   &  2.0164    &  3.7404  &    3.0745  \\
 CN$_1$         &        1.1513   &  1.1473    &  1.0425  &    1.0384  \\
 CN$_2$         &        1.2749   &  1.2720    &  1.1410  &    1.1326  \\
\end{tabular} 
\end{center}
\end{table}

\end{document}